\documentclass[a4paper,12pt]{article}

\usepackage{fancyhdr}
\usepackage{amssymb}
\usepackage{tikz}
\usepackage{graphicx}
\usepackage[colorlinks,hyperfootnotes=false]{hyperref}

\definecolor{yellow text}{rgb}{0.8,0.8,0}

\newcommand{\axes}[2]{
\draw[->] (\xmin,\ymin) -- (\xmax,\ymin) -- ++ (10pt,0pt) node[right]{$#1$};
\draw[->] (\xmin,\ymin) -- (\xmin,\ymax) -- ++ (0pt,10pt) node[right]{$#2$};
}
\newcommand{\helplines}[3][]{
\foreach \x in {#2} \draw[help lines,#1] (\x,\ymin) -- (\x,\ymax);
\foreach \y in {#3} \draw[help lines,#1] (\xmin,\y) -- (\xmax,\y);
}
\newcommand{\axislabels}[2]{
\foreach \x/\xlabel in {#1} \draw (\x,\ymin) -- ++(0pt,-2pt) node[below]{$\xlabel$};
\foreach \y/\ylabel in {#2} \draw (\xmin,\y) -- ++(-2pt,0pt) node[left]{$\ylabel$};
}
\newcommand{\loghelplines}[3][]{
\foreach \log in {-0.6990,-0.5229,-0.3979,-0.3010,-0.2218,-0.1549,-0.0969,-0.0458,0}{
\foreach \x in {#2} \draw[help lines,#1] (\x+\log,\ymin) -- (\x+\log,\ymax);
\foreach \y in {#3} \draw[help lines,#1] (\xmin,\y+\log) -- (\xmax,\y+\log);
}}
\newcommand{\logaxislabels}[2]{
\foreach \x/\xlabel in {#1} \draw (\x,\ymin) -- ++(0pt,-2pt) node[below]{$10^{\xlabel}$};
\foreach \y/\ylabel in {#2} \draw (\xmin,\y) -- ++(-2pt,0pt) node[left]{$10^{\ylabel}$};
}

\renewcommand{\d}[1]{\mathinner{d#1}}
\newcommand{\fn}[2]{\mathinner{#1\mathopen{\left(#2\right)}}}
\newcommand{\order}[1]{\mathinner{\mathcal{O}\mathopen{\left(#1\right)}}}

\newcommand{\eq}[1]{Eq.~(\ref{#1})}
\newcommand{\eqs}[2]{Eqs.~(\ref{#1}) and (\ref{#2})}
\newcommand{\eqss}[3]{Eqs.~(\ref{#1}), (\ref{#2}) and (\ref{#3})}

\newcommand{\eV}{\mathinner{\mathrm{eV}}}

\newcommand{\MeV}{\mathinner{\mathrm{MeV}}}
\newcommand{\GeV}{\mathinner{\mathrm{GeV}}}
\newcommand{\TeV}{\mathinner{\mathrm{TeV}}}

\newcommand{\axino}{{\tilde{a}}}
\newcommand{\nlsp}{N}

\begin{document}

\title{Thermal inflation, baryogenesis and axions}

\author{
Seongcheol Kim, Wan-Il Park and Ewan D. Stewart \\[2ex]
\textit{Department of Physics, KAIST, Daejeon 305-701, South Korea} \\
}

\maketitle

\begin{abstract}
In a previous paper, we proposed a simple extension of the Minimal Supersymmetric Standard Model which gives rise to thermal inflation and baryogenesis in a natural and remarkably consistent way. In this paper, we consider the $\lambda_\phi = 0$ special case of our model, which is the minimal way to incorporate a Peccei-Quinn symmetry. The axino/flatino becomes the lightest supersymmetric particle with $m_\axino \sim 1$ to $10 \GeV$ and is typically over-produced during the flaton decay. Interestingly though, the dark matter abundance is minimized for $m_\axino \sim 1 \GeV$, $f_a \sim 10^{11}$ to $10^{12} \GeV$ and $|\mu| \sim 400 \GeV$ to $2 \TeV$ at an abundance coincident with the observed abundance and with significant amounts of both axions and axinos. Futhermore, for these values the baryon abundance naturally matches the observed abundance.
\end{abstract}

\thispagestyle{fancy}
\rhead{KAIST-TH 2008/05}

\newpage

\section{Introduction}
\label{intro}

In this introduction, we review some necessary background material on axions \cite{Kim:1986ax,KT,Peccei:2006as,Sikivie:2006ni} and thermal inflation \cite{Lyth:1995hj,Lyth:1995ka,Jeong:2004hy,Felder:2007iz}, and then give our motivation for the model presented in this paper.
In Section~\ref{model}, we present our model and describe some of its important features, including a detailed analysis of the flaton decay.
Some technical calculations of flaton decay rates used in this section are given in Appendix~\ref{appendix}.
In Section~\ref{constraints}, we determine the constraints on the model.
In Section~\ref{numerical}, we describe our numerical simulation of the leptogenesis and discuss the results.
In Section~\ref{conclusion}, we summarize the physics of our model, compare it with related models, and discuss observational tests and signatures.

\subsection{Axions}
\label{intax}

The main motivation for axions \cite{Peccei:1977hh,Peccei:1977ur,Weinberg:1977ma,Wilczek:1977pj} \cite{Kim:1979if,Shifman:1979if,Zhitnitsky:1980tq,Dine:1981rt} is to solve the strong $CP$ problem \cite{Kim:1986ax,Peccei:2006as}, though they are also one of the best candidates for dark matter \cite{KT,Sikivie:2006ni}.

The strong $CP$ problem arises because the QCD Lagrangian contains a term
\begin{equation} \label{QCDCP}
\mathcal{L}_\theta = \theta \frac{g^2}{32 \pi^2} F \tilde{F}
\end{equation}
which violates $CP$, and observations require $\theta \lesssim 10^{-10}$ \cite{Dress:1976bq,Altarev:1996xs} which is much smaller than would be expected naively.
To solve the strong $CP$ problem, Peccei and Quinn introduced a Peccei-Quinn (PQ) symmetry, $\mathrm{U}(1)_\mathrm{PQ}$, which is spontaneously broken by the PQ field
\begin{equation} \label{flatondecomp}
\phi = \left( \phi_0 + \frac{\delta r}{\sqrt{2}\,} \right) \exp \left( \frac{i a}{\sqrt{2}\, \phi_0} \right)
\end{equation}
where the pseudo-Nambu-Goldstone boson $a$ is the axion.
A $\mathrm{U}(1)_\mathrm{PQ}$-QCD anomaly adds an extra contribution to \eq{QCDCP} giving
\begin{equation}
\mathcal{L}_\theta = \left( \theta - \frac{a}{f_a} \right) \frac{g^2}{32 \pi^2} F \tilde{F}
\end{equation}
and, after the QCD phase transition, the axion relaxes to cancel $\theta$, solving the strong $CP$ problem.

The axion decay constant $f_a$ is related to the PQ symmetry breaking scale by
\begin{equation} \label{faphi0}
f_a = \frac{\sqrt{2}\, \phi_0}{N}
\end{equation}
$N$ is the coefficient of the $\mathrm{U}(1)_\mathrm{PQ}$-QCD anomaly and is given by
\begin{equation} \label{N}
N = \sum_{i \in \{\mathrm{quarks}\}} p_i
\end{equation}
where the $p_i$ are the PQ charges normalised such that $\phi$ has PQ charge one.
The axion also couples to electromagnetism, indirectly via the $\mathrm{U}(1)_\mathrm{PQ}$-QCD anomaly and directly via the $\mathrm{U}(1)_\mathrm{PQ}$-QED anomaly.
They generate a term in the QED Lagrangian
\begin{equation}
\left( C - \frac{E}{N} \right) \frac{a}{f_a} \frac{e^2}{32 \pi^2} F \tilde{F}
\end{equation}
where $C \simeq 2$ \cite{Buckley:2007tm} and $E$ is the coefficient of the $\mathrm{U}(1)_\mathrm{PQ}$-QED anomaly
\begin{equation} \label{E}
E = 2 \sum_i p_i q_i^2
\end{equation}
where the $q_i$ are the electromagnetic charges.
The axion mass depends on the temperature \cite{KT}
\begin{equation} \label{axionmass}
\fn{m_a}{T} \simeq \left\{
\begin{array}{ccl}
\displaystyle
0.1 \fn{m_a}{0} \left( \frac{\Lambda_\mathrm{QCD}}{T} \right)^{3.7}
& \textrm{for} & \displaystyle
T \gg \Lambda_\mathrm{QCD}
\\[3ex]
\displaystyle
\fn{m_a}{0}
& \textrm{for} & \displaystyle
T \ll \Lambda_\mathrm{QCD}
\end{array}
\right.
\end{equation}
where
\begin{equation} \label{ma0}
\fn{m_a}{0} \simeq 6 \times 10^{-5} \eV \left( \frac{10^{11}\GeV}{f_a} \right)
\end{equation}
and $\Lambda_\mathrm{QCD} \simeq 200 \MeV$ is the scale of the QCD phase transition.

Only invisible axion models with $f_a$ much greater than the electroweak symmetry breaking scale are consistent with observations \cite{Raffelt:2006cw}.
These models can be classified into two types: KSVZ \cite{Kim:1979if,Shifman:1979if} and DFSZ \cite{Zhitnitsky:1980tq,Dine:1981rt}.
In the KSVZ model, the superpotential has a coupling
\begin{equation}
W = \lambda_\chi \phi \chi \bar\chi
\end{equation}
of the PQ field $\phi$ to new quarks $\chi$ and $\bar\chi$ which become heavy after PQ symmetry breaking.
In the DFSZ model, the superpotential has a coupling
\begin{equation}
W = \lambda_\mu \phi^2 H_u H_d
\end{equation}
between the PQ field and the Minimal Supersymmetric Standard Model (MSSM) Higgs fields.

Various astrophysical and cosmological observations constrain $f_a$.
Energy loss in Supernova 1987A gives a lower bound \cite{Raffelt:2006cw}
\begin{equation} \label{facon}
f_a \gtrsim 10^{9} \GeV
\end{equation}
while axion cold dark matter abundance gives an upper bound.
Axion cold dark matter is generated when the axion starts to oscillate coherently at the QCD phase transition due to an initial misalignment of the axion field, and also by the decay of PQ strings formed after PQ symmetry breaking.
Assuming a randomized misalignment angle, as would be expected after a PQ phase transition, misalignment generates \cite{Yao:2006px}
\begin{equation} \label{fama}
\Omega_a \sim 0.2 \left( \frac{f_a}{10^{11} \GeV} \right)^{1.175}
\end{equation}
while the decay of PQ strings generates \cite{Yao:2006px,Battye:1994au,Hagmann:2000ja}
\begin{equation} \label{fasa}
\Omega_a \sim 0.2 \left( \frac{f_a}{10^{10} \textrm{ to } 10^{11} \GeV} \right)^{1.175}
\end{equation}
compared with an observed cold dark matter abundance of \cite{Yao:2006px}
\begin{equation} \label{cdm}
\Omega_\mathrm{CDM} \simeq 0.2
\end{equation}
However, if entropy is released after the QCD phase transition then the axions will be diluted \cite{Yamamoto:1985mb}, see Section~\ref{axcdm}.

The ADMX experiment \cite{admx} has searched for cold dark matter axions with axion mass in the range $ m_a \simeq \left( 2 \textrm{ to } 3 \right) \times 10^{-6}\eV$ corresponding to $f_a \sim \left( 2 \textrm{ to } 3 \right) \times10^{12} \GeV$.
For this range of axion masses, the resulting constraint on the ratio of the axion-photon coupling to axion mass translates to \cite{Buckley:2007tm}
\begin{equation}
0 \textrm{ to } 0.8 \lesssim \frac{E}{N} \lesssim 3.6 \textrm{ to } 4
\end{equation}

\subsection{Thermal inflation}
\label{introti}

The main motivation for thermal inflation \cite{Lyth:1995hj,Lyth:1995ka} \cite{Yamamoto:1985mb,Yamamoto:1985rd,Enqvist:1985kz,Bertolami:1987xb,Ellis:1986nn,Ellis:1989ii} \cite{Randall:1994fr} is to solve the moduli problem \cite{Coughlan:1983ci,Banks:1993en,de Carlos:1993jw}, though it also solves the gravitino problem \cite{Khlopov:1984pf,Ellis:1984eq} and may be able to explain the origin of the observed baryon asymmetry \cite{Jeong:2004hy,Felder:2007iz} \cite{Stewart:1996ai,Kawasaki:2006py} \cite{Lazarides:1985ja,Yamamoto:1986jw,Mohapatra:1986dg,Lazarides:1987yq}.
Irrespective of these uses, thermal inflation is sufficiently natural that it might be expected to occur anyway.

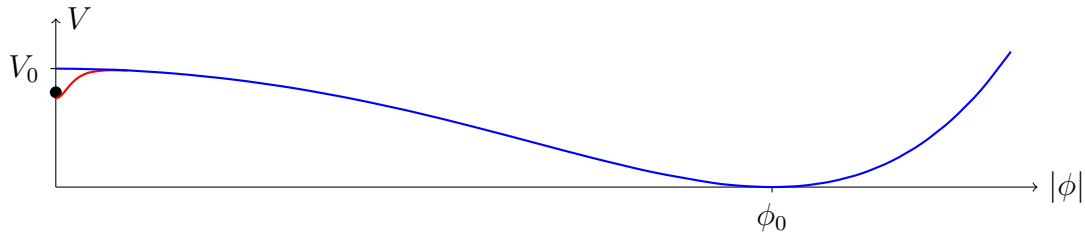
\begin{figure}[b]
\centering
\newcommand{\xmin}{0}
\newcommand{\xmax}{8}
\newcommand{\ymin}{0}
\newcommand{\ymax}{1.2}
\begin{tikzpicture}[x=0.1\textwidth,y=0.1\textwidth]
\axes{|\phi|}{V}
\axislabels{6/\phi_0}{1/V_0}
\fill (0,0.8) circle(0.05);
\draw[red,thick] plot[smooth] coordinates { (0,0.75) (0.025,0.758) (0.05,0.778) (0.075,0.807) (0.1,0.838) (0.125,0.867) (0.15,0.893) (0.175,0.914) (0.2,0.932) (0.25,0.957) (0.3,0.971) (0.4,0.984) (0.5,0.986) (0.6,0.984) };
\draw[blue,thick] plot[smooth] coordinates { (0,1) (0.5,0.990) (1,0.958) (1.5,0.906) (2,0.834) (2.5,0.742) (3,0.633) (3.5,0.509) (4,0.377) (4.5,0.245) (5,0.126) (5.5,0.036) (5.75,0.010) (6,0) (6.25,0.011) (6.5,0.048) (6.75,0.115) (7,0.219) (7.25,0.366) (7.5,0.564) (7.75,0.819) (8,1.143) };
\end{tikzpicture}
\caption{ \label{fig:tip}
Thermal inflation occurs when a flaton $\phi$ is held at the origin by its finite temperature potential and $V_0$ dominates the energy density.
}
\end{figure}

A flaton is a scalar field $\phi$ with negative mass squared at the origin and no quartic term
\begin{equation} \label{fpidea}
V(\phi) = V_0 - m_\phi^2 |\phi|^2 + \ldots
\end{equation}
so that its vacuum expectation value $\phi_0$ is much greater than its mass scale
\begin{equation}
\phi_0 \gg m_\phi
\end{equation}
with $m_\phi$ a typical soft supersymmetry breaking mass
\begin{equation} \label{mphirange}
m_\phi \sim m_\mathrm{s} \sim 10^2 \textrm{ to } 10^3 \GeV
\end{equation}
See Figure~\ref{fig:tip}.

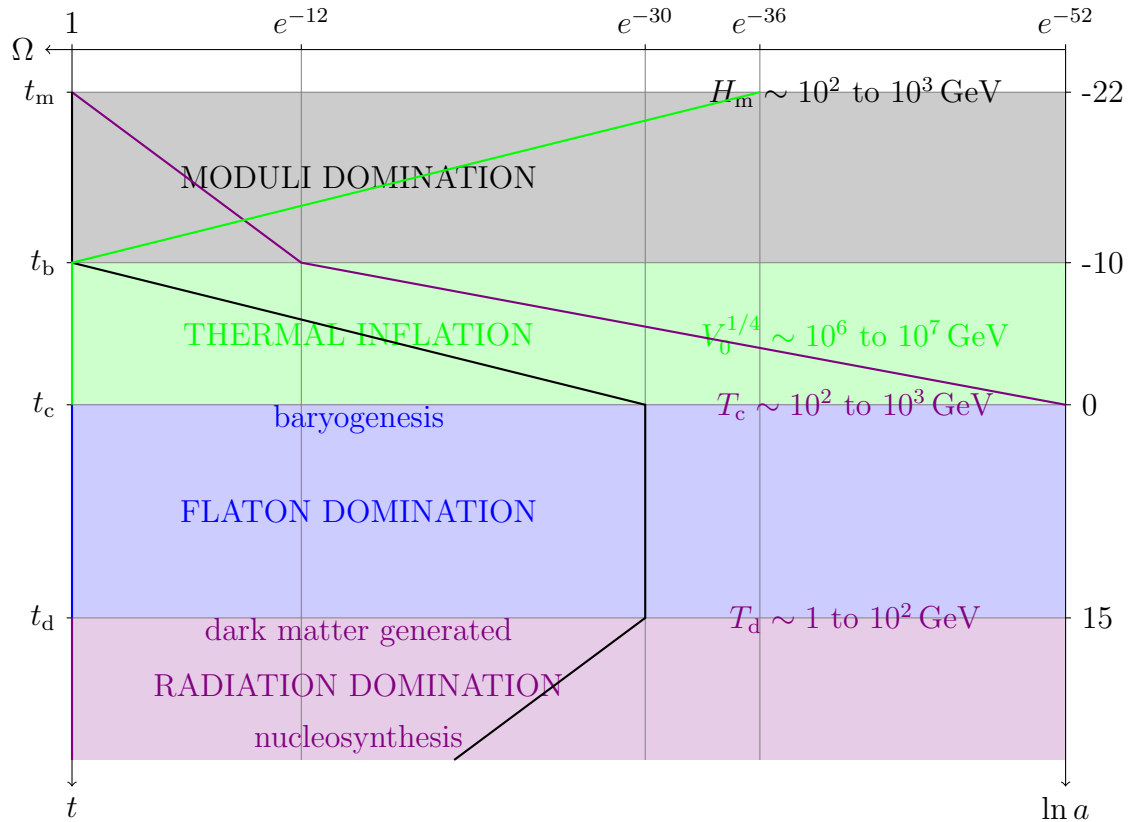
\begin{figure}[t]
\centering
\newcommand{\xmin}{0}
\newcommand{\xmax}{52}
\newcommand{\ymin}{-25}
\newcommand{\ymax}{25}
\begin{tikzpicture}[x=0.016\textwidth,y=0.012\textwidth]
\fill[black!20] (\xmin,22) rectangle (\xmax,10);
\fill[green!20] (\xmin,10) rectangle (\xmax,0);
\fill[blue!20] (\xmin,0) rectangle (\xmax,-15);
\fill[violet!20] (\xmin,-15) rectangle (\xmax,-25);
\helplines{12,30,36}{-15,0,10,22}
\draw[->] (\xmax,\ymax) -- (\xmin,\ymax) -- ++ (-10pt,0pt) node[left]{$\Omega$};
\draw[->] (\xmin,\ymax) -- (\xmin,\ymin) -- ++ (0pt,-10pt) node[below]{$t$};
\draw[->] (\xmax,\ymax) -- (\xmax,\ymin) -- ++ (0pt,-10pt) node[below]{$\ln a$};
\draw (0,\ymax) -- ++(0pt,2pt) node[above]{1};
\draw (12,\ymax) -- ++(0pt,2pt) node[above]{$e^{-12}$};
\draw (30,\ymax) -- ++(0pt,2pt) node[above]{$e^{-30}$};
\draw (36,\ymax) -- ++(0pt,2pt) node[above]{$e^{-36}$};
\draw (52,\ymax) -- ++(0pt,2pt) node[above]{$e^{-52}$};
\draw (\xmin,22) -- ++(-2pt,0pt) node[left]{$t_\mathrm{m}$};
\draw (\xmin,10) -- ++(-2pt,0pt) node[left]{$t_\mathrm{b}$};
\draw (\xmin,0) -- ++(-2pt,0pt) node[left]{$t_\mathrm{c}$};
\draw (\xmin,-15) -- ++(-2pt,0pt) node[left]{$t_\mathrm{d}$};
\draw (\xmax,22) -- ++(2pt,0pt) node[right]{-22};
\draw (\xmax,10) -- ++(2pt,0pt) node[right]{-10};
\draw (\xmax,0) -- ++(2pt,0pt) node[right]{0};
\draw (\xmax,-15) -- ++(2pt,0pt) node[right]{15};
\node[black] at (41,22) {$H_\mathrm{m} \sim 10^2 \textrm{ to } 10^3 \GeV$};
\node[black] at (15,16) {MODULI DOMINATION};
\node[green] at (15,5) {THERMAL INFLATION};
\node[green] at (41,5) {$V_0^{1/4} \sim 10^6 \textrm{ to } 10^7 \GeV$};
\node[violet] at (41,0) {$T_\mathrm{c} \sim 10^2 \textrm{ to } 10^3 \GeV$};
\node[blue] at (15,-1) {baryogenesis};
\node[blue] at (15,-7.5) {FLATON DOMINATION};
\node[violet] at (41,-15) {$T_\mathrm{d} \sim 1 \textrm{ to } 10^2 \GeV$};
\node[violet] at (15,-16) {dark matter generated};
\node[violet] at (15,-19.75) {RADIATION DOMINATION};
\node[violet] at (15,-23.5) {nucleosynthesis};
\draw[violet,thick] (0,22) -- (12,10) -- (52,0);
\draw[black,thick] (0,22) -- (0,10) -- (30,0) -- (30,-15) -- (20,-25);
\draw[green,thick] (36,22) -- (0,10) -- (0,0);
\draw[blue,thick] (0,0) -- (0,-15);
\draw[violet,thick] (0,-15) -- (0,-25);
\end{tikzpicture}
\caption{ \label{fig:tih}
History of the universe with thermal inflation.
The fractional density $\Omega_X \equiv \rho_X/\rho$ of moduli, {\color{green} potential}, {\color{blue} flaton} and {\color{violet} radiation} is plotted against the number of $e$-folds of expansion $\ln a$.
}
\end{figure}

Thermal inflation begins when the flaton is held at the origin by its finite temperature potential and the energy density has dropped sufficiently for the potential energy at the origin to dominate. It lasts for about 10 $e$-folds and ends in a first order phase transition just before the temperature drops to the critical temperature
\begin{equation}
T_\mathrm{c} \sim m_\phi
\end{equation}
for the flaton to roll away from the origin. It is followed by a period of flaton matter domination until the flaton decays leaving a radiation dominated universe at temperature $T_\mathrm{d}$. See Figure~\ref{fig:tih}.

\subsubsection{Thermal inflation and the moduli problem}

Moduli are scalar fields with Planckian vacuum expectation values, and hence gravitational strength interactions. Their potential arises due to supersymmetry breaking, and, assuming supersymmetry breaking is transmitted to the observable sector via gravitational strength interactions, they would be expected to have vacuum masses of order the soft supersymmetry breaking scale in the observable sector
\begin{equation}
m_\mathrm{mod} \sim m_\mathrm{s}
\end{equation}

However, in the early universe, the finite energy density breaks supersymmetry. When $H \gtrsim m_\mathrm{s}$ this supersymmetry breaking dominates over the vacuum supersymmetry breaking and hence determines the moduli potential. When $H$ drops below $m_\mathrm{s}$ the moduli potential reduces to its vacuum form, but with the moduli typically displaced by a Planckian distance. The moduli then start oscillating with Planckian amplitude and immediately dominate the energy density of the universe, and, because of their relatively low mass and very weak interactions, they persist beyond nucleosynthesis with disastrous consequences \cite{Coughlan:1983ci,Banks:1993en,de Carlos:1993jw}.

Inflation is typically invoked to rid the universe of unwanted relics, but one would expect an inflaton to have a mass $\gtrsim m_\mathrm{s}$ and hence primordial inflation to occur at a scale $H \gtrsim m_\mathrm{s}$. However, the moduli are generated at $H \sim m_\mathrm{s}$, and to a lesser extent by any phase transition at $H \lesssim m_\mathrm{s}$. Thus one wants inflation at $H \ll m_\mathrm{s}$ to dilute the moduli, but it is very difficult to realize primordial inflation at these scales.

Thermal inflation \cite{Lyth:1995hj,Lyth:1995ka} on the other hand automatically occurs at $H \ll m_\mathrm{s}$. For a thermal inflation scale
\begin{equation}
V_0^{1/4} \sim 10^6 \textrm{ to } 10^7 \GeV
\end{equation}
corresponding to
\begin{equation} \label{phi0range}
\phi_0 \sim 10^{10} \textrm{ to } 10^{12} \GeV
\end{equation}
thermal inflation provides enough dilution to rid the universe of moduli, but has a low enough scale not to regenerate them afterwards. Furthermore, it does not last long, only around 10 $e$-folds, and so does not destroy the primordial perturbations needed for structure formation, only shifting them to slightly larger scales.

In more detail, following Ref.~\cite{Lyth:1995ka}, the moduli produced before thermal inflation are diluted to an abundance
\begin{equation}
\frac{n_\mathrm{mod}}{s} \sim \frac{T_\mathrm{c}^3 T_\mathrm{d}}{m_\mathrm{mod}^{1/2} V_0}
\end{equation}
and this is reduced further in the case of double thermal inflation, and those generated at the end of thermal inflation have an abundance
\begin{equation} \label{modulicon}
\frac{n_\mathrm{mod}}{s} \sim \frac{V_0 T_\mathrm{d}}{m_\mathrm{mod}^3}
\end{equation}
with nucleosynthesis requiring \cite{Kawasaki:2004qu}
\begin{equation} \label{bbnmodbd}
\frac{n_\mathrm{mod}}{s} \lesssim 10^{-12}
\end{equation}

\subsubsection{Thermal inflation and baryogenesis}
\label{introbaryo}

Thermal inflation provides a very natural solution to the moduli problem, but unfortunately is incompatible with most baryogenesis scenarios since it dilutes any baryons generated before it and the temperature after flaton decay is very low, typically $\order{10\GeV}$. Fortunately, it gives rise to its own baryogenesis scenario \cite{Jeong:2004hy,Felder:2007iz} \cite{Stewart:1996ai,Kawasaki:2006py}.

In Ref.~\cite{Jeong:2004hy}, we proposed the simple extension of the MSSM
\begin{equation} \label{Woriginal}
W = \lambda_u Q H_u \bar{u} + \lambda_d Q H_d \bar{d} + \lambda_e L H_d \bar{e}
+ \lambda_\mu \phi^2 H_u H_d + \frac{1}{2} \lambda_\nu \left( L H_u \right)^2
+ \lambda_\chi \phi \chi \bar{\chi} + \frac{1}{4} \lambda_\phi \phi^4
\end{equation}
with the key extra assumption that the soft supersymmetry breaking mass squared along the $LH_u$ flat direction is negative
\begin{equation} \label{key}
- m_{LH_u}^2 = \frac{1}{2} \left( m^2_L - m^2_{H_u} \right) < 0
\end{equation}
where $m^2_L$ and $-m^2_{H_u}$ are the soft supersymmetry breaking mass squareds of $L$ and $H_u$.
The first three terms in \eq{Woriginal} are MSSM terms and the fifth gives neutrino masses.
The term $\lambda_\chi \phi \chi \bar{\chi}$ couples the flaton $\phi$ to the thermal bath.
After thermal inflation, the $\chi$ and $\bar\chi$ fields acquire large masses
\begin{equation}
M_\chi = \lambda_\chi \phi_0
\end{equation}
and so are not strongly constrained apart from that they should not damage gauge coupling unification.
This coupling also induces $|\phi|$ dependent renormalization group running of $\phi$'s effective soft supersymmetry breaking mass squared, tending to drive it negative at small $|\phi|$, as is required for a flaton.
The term $\frac{1}{4} \lambda_\phi \phi^4$ stabilizes $\phi$'s potential at a value
\begin{equation}
\phi_0 \sim \sqrt{ \frac{m_\phi}{\lambda_\phi} }
\end{equation}
that turns $\lambda_\mu \phi^2 H_u H_d$ into the MSSM $\mu$-term with
\begin{equation}
|\mu| = \left| \lambda_\mu \phi_0^2 \right| \sim \left| \frac{\lambda_\mu}{\lambda_\phi} \right| m_\phi
\end{equation}
The $\lambda_\mu \phi^2 H_u H_d$ coupling also helps to ensure that the temperature after flaton decay $T_\mathrm{d}$ is high enough for dark matter to be generated.
Thus we have a simple model of thermal inflation with the right scale for $\phi_0$ if $\lambda_\phi \sim M_\mathrm{Pl}^{-1}$, and that also generates a $\mu$-term of the right scale if $\lambda_\mu \sim \lambda_\phi$.

Our key assumption, \eq{key}, seems at first sight dangerous since it implies a deep non-MSSM vacuum with $LH_u \sim (10^9 \GeV)^2$ and
\begin{equation} \label{nmssm}
\lambda_d Q L \bar{d} + \lambda_e L L \bar{e} = - \mu L H_u
\end{equation}
eliminating the $\mu$-term contribution to $LH_u$'s mass squared \cite{Casas:1995pd}.
To analyze the dynamics induced by \eq{key}, we parameterize the potentially unstable flat directions as
\begin{equation}
L = \left(\begin{array}{c}
e/\sqrt{2} \\ l
\end{array}\right)
\quad , \quad
H_u = \left(\begin{array}{c}
h_u \\ 0
\end{array}\right)
\quad , \quad
H_d = \left(\begin{array}{c}
0 \\ h_d
\end{array}\right)
\quad , \quad
\bar{e} = \left(\begin{array}{ccc}
e/\sqrt{2}
\end{array}\right)
\end{equation}
\begin{equation}
\bar{u} = \left(\begin{array}{ccc}
0 & 0 & 0 \\
\end{array}\right)
\quad , \quad
Q = \left(\begin{array}{ccc}
d/\sqrt{2} & 0 & 0 \\ 0 & 0 & 0
\end{array}\right)
\quad , \quad
\bar{d} = \left(\begin{array}{ccc}
d/\sqrt{2} & 0 & 0 \\
\end{array}\right)
\end{equation}
\begin{equation}
\phantom{L = \left(\begin{array}{c} e/\sqrt{2} \\ l \end{array}\right)}
\phi = \phi
\quad , \quad
\chi = 0
\quad , \quad
\bar\chi = 0
\phantom{L = \left(\begin{array}{c} e/\sqrt{2} \\ l \end{array}\right)}
\end{equation}
The superpotential \eq{Woriginal} reduces to
\begin{equation}
W = \frac{1}{2} \lambda_d h_d d^2 + \frac{1}{2} \lambda_e h_d e^2 + \lambda_\mu \phi^2 h_u h_d + \frac{1}{2} \lambda_\nu \left( l h_u \right)^2 + \frac{1}{4} \lambda_\phi \phi^4
\end{equation}
with the remaining $D$-term constraint
\begin{equation}
D = |h_u|^2 - |h_d|^2 - |l|^2 + \frac{1}{2} |d|^2 + \frac{1}{2} |e|^2 = 0
\end{equation}
The potential is
\begin{eqnarray}
V & = & V_0 - m_\phi^2 |\phi|^2 + m_L^2 |l|^2 - m_{H_u}^2 |h_u|^2 + m_{H_d}^2 |h_d|^2 + m^2_d |d|^2 + m^2_e |e|^2
\nonumber \\ && {}
+ \left( \frac{1}{2} A_\nu \lambda_\nu l^2 h_u^2
+ A_\mu \lambda_\mu \phi^2 h_u h_d
+ \frac{1}{4} A_\phi \lambda_\phi \phi^4
+ \frac{1}{2} A_d \lambda_d h_d d^2 + \frac{1}{2} A_e \lambda_e h_d e^2 + \textrm{c.c.} \right)
\nonumber \\ && {}
+ \left| \lambda_\nu l h_u^2 \right|^2
+ \left| \lambda_\nu l^2 h_u + \lambda_\mu \phi^2 h_d \right|^2
+ \left| \lambda_\phi \phi^3 + 2 \lambda_\mu \phi h_u h_d \right|^2
+ \left| \lambda_d h_d d \right|^2 + \left| \lambda_e h_d e \right|^2
\nonumber \\ \label{Voriginal} && {}
+ \left| \lambda_\mu \phi^2 h_u + \frac{1}{2} \lambda_d d^2 + \frac{1}{2} \lambda_e e^2 \right|^2
+ \frac{1}{2} g^2 \left( |h_u|^2 - |h_d|^2 - |l|^2 + \frac{1}{2} |d|^2 + \frac{1}{2} |e|^2 \right)^2
\end{eqnarray}

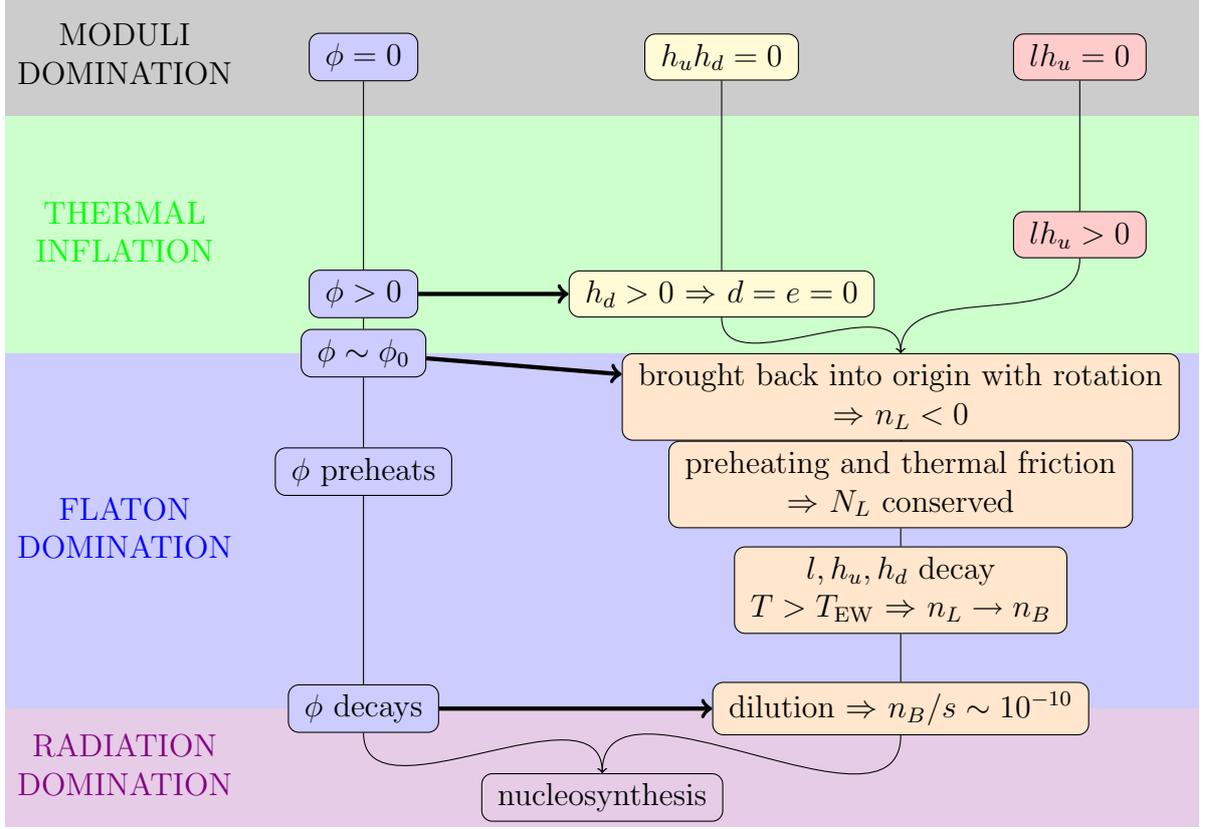
\begin{figure}[t]
\centering
\newcommand{\bubble}{\node[draw,rectangle,rounded corners,inner xsep=\tabcolsep,inner ysep=0.25\baselineskip]}
\newcommand{\bubblearray}{\node[draw,rectangle,rounded corners,inner xsep=0pt,inner ysep=0.125\baselineskip]}
\begin{tikzpicture}[x=0.1\textwidth,y=0.1\textwidth]
\fill[black!20] (-5,6) rectangle (5,7);
\node[black] at (-4,6.5) {\begin{tabular}{c} MODULI \\ DOMINATION \end{tabular}};
\fill[green!20] (-5,4) rectangle (5,6);
\node[green] at (-4,5) {\begin{tabular}{c} THERMAL \\ INFLATION \end{tabular}};
\fill[blue!20] (-5,1) rectangle (5,4);
\node[blue] at (-4,2.5) {\begin{tabular}{c} FLATON \\ DOMINATION \end{tabular}};
\fill[violet!20] (-5,0) rectangle (5,1);
\node[violet] at (-4,0.5) {\begin{tabular}{c} RADIATION \\ DOMINATION \end{tabular}};
\bubble[fill=blue!20](f8) at (-2,6.5) {$\phi = 0$};
\bubble[fill=blue!20](f6) at (-2,4.5) {$\phi > 0$};
\bubble[fill=blue!20](f5) at (-2,4) {$\phi \sim \phi_0$};
\bubble[fill=blue!20](f4) at (-2,3) {$\phi$ preheats};
\bubble[fill=blue!20](f2) at (-2,1) {$\phi$ decays};
\bubble[fill=yellow!20](d8) at (1,6.5) {$h_uh_d=0$};
\bubble[fill=yellow!20](d6) at (1,4.5) {$h_d > 0$ $\Rightarrow$ $d = e = 0$};
\bubble[fill=red!20](l8) at (4,6.5) {$lh_u=0$};
\bubble[fill=red!20](l7) at (4,5) {$lh_u>0$};
\bubblearray[below,fill=orange!20](a5) at (2.5,4) {\begin{tabular}{c} brought back into origin with rotation \\ $\Rightarrow$ $n_L < 0$ \end{tabular}};
\bubblearray[below,fill=orange!20](a4) at (a5.south) {\begin{tabular}{c} preheating and thermal friction \\ $\Rightarrow$ $N_L$ conserved \end{tabular}};
\bubblearray[fill=orange!20](a3) at (2.5,2) {\begin{tabular}{c} $l, h_u, h_d$ decay \\ $T > T_\mathrm{EW}$ $\Rightarrow$ $n_L \rightarrow n_B$ \end{tabular}};
\bubble[fill=orange!20](a2) at (2.5,1) {dilution $\Rightarrow$ $n_B/s \sim 10^{-10}$};
\bubble[fill=violet!20](r1) at (0,0.25) {nucleosynthesis};
\draw[->] (f8) -- (f6) -- (f5) -- (f4) -- (f2) to[out=270,in=90] (r1);
\draw[->] (d8) -- (d6) to[out=270,in=90] (a5);
\draw[->] (l8) -- (l7) to[out=270,in=90] (a5);
\draw[->] (a5) -- (a4) -- (a3) -- (a2) to[out=270,in=90] (r1);
\draw[->,ultra thick] (f6) -- (d6);
\draw[->,ultra thick] (f5) -- (a5);
\draw[->,ultra thick] (f2) -- (a2);
\end{tikzpicture}
\caption{ \label{fig:tib}
Thermal inflation baryogenesis.
}
\end{figure}

The dynamics is illustrated in Figure~\ref{fig:tib}. Initially, all fields are held at the origin by their finite temperature potential. Once the energy density has dropped sufficiently, $V_0$ will dominate driving thermal inflation. After the temperature drops further, one of the unstable directions, $\phi$ or $lh_u$, will roll away from the origin. We assume $lh_u$ rolls away first. Then $\frac{1}{2} A_\nu \lambda_\nu l^2 h_u^2$ fixes the phase of $lh_u$, and $\left| \lambda_\nu l h_u^2 \right|^2$ and $\left| \lambda_\nu l^2 h_u \right|^2$ stabilize its magnitude. The $lh_u$ field may partially reheat the thermal bath and so prolong the thermal inflation, but eventually $\phi$ will also roll away from the origin, ending thermal inflation. As $\phi$ rolls away, $A_\mu \lambda_\mu \phi^2 h_u h_d$ will force $h_d$ to become non-zero. $\left| \lambda_d h_d d \right|^2$ and $\left| \lambda_e h_d e \right|^2$ then constrain $d$ and $e$ to zero, shielding the dynamics from the dangerous non-MSSM vacuum in the direction of \eq{nmssm}. Then $A_\mu \lambda_\mu \phi^2 h_u h_d$, and also $\frac{1}{4} A_\phi \lambda_\phi \phi^4$ and the cross term from $\left| \lambda_\phi \phi^3 + 2 \lambda_\mu \phi h_u h_d \right|^2$, fix the phase of $\phi^2 h_u h_d$. As $\phi$ nears its minimum, the cross term from $\left| \lambda_\nu l^2 h_u + \lambda_\mu \phi^2 h_d \right|^2$ rotates the phase of $lh_u$ generating a lepton asymmetry, and at the same time $\left| \lambda_\mu \phi^2 h_u \right|^2$ gives an extra contribution to the mass squareds of $lh_u$ and $h_uh_d$, bringing them back in towards the origin. Thus we have a type of Affleck-Dine (AD) leptogenesis \cite{Affleck:1984fy,Dine:1995uk,Dine:1995kz}. Preheating then damps the amplitude of the $lh_u$ and $h_uh_d$ fields keeping them in the lepton preserving region near the origin \cite{Felder:2007iz}. The $lh_u$ and $h_uh_d$ fields then decay, at a temperature in the MSSM sector above the electroweak scale, and their lepton number is converted to baryon number by sphalerons. Finally, the flaton decays, diluting the baryon density to the value required by observations, $n_B/s \sim 10^{-10}$.

\subsubsection{Thermal inflation and axions}

The scale of the flaton vacuum expectation value, \eq{phi0range}, coincides with that of the PQ field, Eqs.~(\ref{faphi0}) and (\ref{facon}) to (\ref{cdm}).
This motivates unification of thermal inflation and axions \cite{Lyth:1995hj,Lyth:1995ka}, and several papers have studied this in detail \cite{Choi:1996vz,Lazarides:2000em,Chun:2000jr,Chun:2000jx}.
In order to introduce a $\mathrm{U}(1)_\mathrm{PQ}$ symmetry, Refs.~\cite{Choi:1996vz,Lazarides:2000em,Chun:2000jr,Chun:2000jx} extended the flaton sector, $\phi \to \phi_1 , \phi_2$, to get a multi-field flaton axion model of the type studied in Refs.~\cite{Lazarides:1985bj,Casas:1987bw,Murayama:1992dj}.
In this type of thermal inflation axion model, the main danger is that too many hot axions will be produced in the flaton decay to be consistent with Big Bang nucleosynthesis.
This imposes significant constraints on the parameter space.

\subsection{This paper}

The motivation for this paper starts from noting that our thermal inflation baryogenesis model, \eqs{Woriginal}{key}, generates a very rich and complex but remarkably consistent cosmology from a very simple and constrained extension of the MSSM.

As discussed in Section~\ref{introbaryo}, the first three terms in \eq{Woriginal} are pure MSSM terms. The fourth term generates the MSSM $\mu$-term, forces $H_d$ to become non-zero shielding the dynamics from the deeper non-MSSM vacuum, in combination with the fifth term generates the lepton asymmetry, couples the flaton to $LH_u$ and $H_uH_d$ allowing the flaton to bring them back in towards the origin, and provides an efficient decay channel for the flaton allowing the temperature after flaton decay to be high enough for dark matter generation. The fifth term generates neutrino masses, stabilizes the $LH_u$ field during thermal inflation, and in combination with the fourth term generates the lepton asymmetry. The sixth term couples the flaton to the thermal bath holding it at the origin during thermal inflation, and drives the flaton's mass squared negative at small $|\phi|$. The seventh and last term stabilizes the flaton field.

Thus, all the terms we add, though simple and natural, do \emph{many} different things. However, there is one exception. The term $\frac{1}{4} \lambda_\phi \phi^4$ is simply introduced to stabilize the flaton's potential at the right scale. If we eliminate this term, the flaton's potential will in any case be stabilized by the $|\phi|$ dependent renormalization of the flaton's mass squared induced by the coupling $\lambda_\chi \phi \chi \bar{\chi}$. Furthermore, the model now has a $\mathrm{U}(1)_\mathrm{PQ}$ symmetry and so automatically includes the axion without any need to introduce extra fields. This simplest of axion models was studied long ago by Moxhay and Yamamoto \cite{Moxhay:1984am} but has since been largely ignored. Its one disadvantage is that we have to tune $\phi_0$ to the right scale, though this tuning is very modest.

Thus we have an even simpler\footnote{It is arguable whether the absence of the $\phi^4$ term is more natural or not.} and more constrained extension of the MSSM, \eq{Woriginal} with $\lambda_\phi = 0$, that generates an even richer and more complex, but as we will see, even more remarkably consistent cosmology.

\section{Model}
\label{model}

Our model \footnote{An obvious alternative would be to replace $\lambda_\mu \phi^2 H_u H_d$ and $\lambda_\mu \sim M_\mathrm{Pl}^{-1}$ with $\lambda_\mu \phi^3 H_u H_d$ and $\lambda_\mu \sim M_\mathrm{GUT}^{-2}$.}
\begin{equation} \label{W}
W = \lambda_u Q H_u \bar{u} + \lambda_d Q H_d \bar{d} + \lambda_e L H_d \bar{e}
+ \lambda_\mu \phi^2 H_u H_d + \frac{1}{2} \lambda_\nu \left( L H_u \right)^2
+ \lambda_\chi \phi \chi \bar\chi
\end{equation}
with the key parameter condition \eq{key}, is a simple extension of the MSSM incorporating thermal inflation, baryogenesis and axions. In this section we describe some of its important properties.

\subsection{Flaton potential}
\label{potential}

The flaton potential has the form
\begin{equation} \label{Vphi}
\fn{V}{\phi} = V_0 - \fn{f}{\frac{1}{2} \alpha_\phi \ln\frac{\left|\phi\right|^2 + m_\mathrm{s}^2}{m_\mathrm{s}^2}} m_\phi^2 |\phi|^2
\end{equation}
where the function $f$ encodes the $|\phi|$ dependent renormalization of the flaton's mass squared.
We set
\begin{equation}
\fn{f}{0} \equiv 1
\end{equation}
to fix the definition of $m_\phi^2$ at the origin and recover \eq{fpidea}
\begin{equation}
\fn{V}{\phi} = V_0 - m_\phi^2 |\phi|^2 + \ldots
\qquad \textrm{for } |\phi| \ll m_\mathrm{s}
\end{equation}
Away from the origin the potential simplifies to
\begin{equation} \label{fpaway}
\fn{V}{\phi} = V_0 - \fn{f}{\alpha_\phi \ln\frac{\left|\phi\right|}{m_\mathrm{s}}} m_\phi^2 |\phi|^2
\qquad \textrm{for } |\phi| \gg m_\mathrm{s}
\end{equation}
Now
\begin{eqnarray} \label{fpp}
\frac{d V}{d|\phi|} & = & - \left( 2 f + \alpha_\phi f' \right) m_\phi^2 |\phi|
\\ \label{fppp}
\frac{d^2 V}{d|\phi|^2} & = & - \left( 2 f + 3 \alpha_\phi f' + \alpha_\phi^2 f'' \right) m_\phi^2
\end{eqnarray}
Therefore, the potential has a minimum at $|\phi| = \phi_0$ with
\begin{equation} \label{f0}
f_0 = - \frac{1}{2} \alpha_\phi f'_0
\end{equation}
We set
\begin{equation} \label{f0p}
f'_0 \equiv -1
\end{equation}
to fix the definition of $\alpha_\phi$ at $|\phi| = \phi_0$.
Setting the vacuum energy to zero, \eqss{fpaway}{f0}{f0p} give
\begin{equation} \label{V0}
V_0 = \frac{1}{2} \alpha_\phi m_\phi^2 \phi_0^2
\end{equation}
and \eqss{fppp}{f0}{f0p} give the physical mass squared at the minimum
\begin{equation} \label{flatonmass}
m_\mathrm{PQ}^2 \equiv \frac{1}{2} \left. \frac{d^2 V}{d|\phi|^2} \right|_{|\phi|=\phi_0}
= \left( 1 - \frac{1}{2} \alpha_\phi f''_0 \right) \alpha_\phi m_\phi^2
\simeq \alpha_\phi m_\phi^2
\end{equation}
Note the suppression factor of $\alpha_\phi$ in both \eqs{V0}{flatonmass}.
In the vicinity of the minimum
\begin{eqnarray}
\fn{V}{\phi} = V_0 - \left[ \frac{1}{2} \alpha_\phi - \alpha_\phi \ln{ \left| \frac{\phi}{\phi_0} \right| } + \order{ \alpha_\phi^2 \ln^2{ \left| \frac{\phi}{\phi_0} \right| } } \right] m_\phi^2 |\phi|^2
\end{eqnarray}

The value of $\phi_0$ is determined by the $|\phi|$ dependent renormalisation group running of the flaton's soft supersymmetry breaking mass squared.
The renormalization coefficient $\alpha_\phi$ is determined by the coupling $\lambda_\chi \phi \chi \bar\chi$ in \eq{W}
\begin{equation} \label{alphaphi}
\alpha_\phi = \frac{1}{8\pi^2 m_\phi^2} \sum_\chi \left| \lambda_\chi \right|^2 \left( m^2_\chi + m^2_{\bar\chi} + \left| A_\chi \right|^2 \right)
\end{equation}
where $m_\chi$, $m_{\bar\chi}$ and $A_\chi$ are the soft supersymmetry breaking masses and $A$ parameter of $\chi$ and $\bar\chi$.
Therefore, using \eq{alphatoy}, at least some of the Yukawa couplings should be unsuppressed
\begin{equation} \label{lambdachiest}
\sum_\chi \left| \lambda_\chi \right|^2 \sim 1
\end{equation}
in order to obtain the correct scale for $\phi_0$.

For definiteness, in our numerical simulation described in Section~\ref{numerical}, we consider the simplest case
\begin{equation} \label{f}
\fn{f}{x} = 1 - x
\end{equation}
We can then solve \eq{f0} to give
\begin{equation} \label{phi0}
\phi_0 = m_\mathrm{s} \exp \left( \frac{1}{\alpha_\phi} - \frac{1}{2} \right)
\end{equation}
To match \eqs{mphirange}{phi0range} then requires
\begin{equation} \label{alphatoy}
\alpha_\phi \simeq 0.05
\end{equation}

\subsection{Lightest supersymmetric particle}
\label{lsp}

In our model, the axino/flatino mass is generated by a $\chi \bar\chi$ loop containing the soft supersymmetry-breaking interaction $A_\chi \lambda_\chi \phi \chi \bar\chi$ \cite{Moxhay:1984am}
\begin{equation} \label{axinomass}
m_\axino \simeq \frac{1}{16 \pi^2} \sum_\chi \lambda_\chi^2 A_\chi
\sim 1 \textrm{ to } 10 \GeV
\end{equation}
where we have used \eq{lambdachiest} to estimate $\lambda_\chi$.
Therefore the axino will be the lightest supersymmetric particle (LSP) in our model.

\subsection{$\chi$ and $\bar\chi$ and the PQ anomalies}
\label{pq}

To maintain gauge coupling unification, $\chi$ and $\bar\chi$ should form complete $\mathrm{SU}(5)$ multiplets, and so, assuming that they are not singlets, they should contain fields charged under $\mathrm{SU}(3)$.
Thus the Peccei-Quinn field $\phi$ has couplings to both heavy quarks, contained in $\chi$ and $\bar\chi$, and the Higgs fields $H_u$ and $H_d$.
Thus our model is a combination of the KSVZ and DFSZ axion models.

Assuming that $\chi$ can be decomposed into $\mathbf{1}$, $\mathbf{5}$ and $\mathbf{10}$ representations of $\mathrm{SU}(5)$, and $\bar\chi$ into corresponding $\mathbf{\bar{1}}$, $\mathbf{\bar{5}}$ and $\mathbf{\bar{10}}$ representations, we can parameterize $\chi$ and $\bar\chi$ as
\begin{eqnarray} \label{chiN}
\chi & = & N_1 \mathbf{1} + N_5 \mathbf{5} + N_{10} \mathbf{10}
\\ \label{chiNbar}
\bar\chi & = & N_1 \mathbf{\bar{1}} + N_5 \mathbf{\bar{5}} + N_{10} \mathbf{\bar{10}}
\end{eqnarray}
with the number of heavy quarks
\begin{equation} \label{nq}
N_q = N_5 + 3 N_{10}
\end{equation}
Then \eq{N} gives
\begin{equation} \label{ourN}
N = 6 - N_q
\end{equation}
with the existence of an axion requiring $N \neq 0$, and \eq{E} gives
\begin{equation} \label{ourE}
E = 12 - \frac{8}{3} N_q
\end{equation}
and therefore
\begin{equation}
\frac{E}{N} = \frac{8}{3} - \frac{4}{N}
\end{equation}

As discussed above, $\chi$ and $\bar\chi$ should form complete representations of $\mathrm{SU}(5)$ in order to preserve gauge coupling unification.
However, although gauge coupling unification is preserved, the GUT scale gauge coupling becomes stronger than in the pure MSSM case.
To preserve \emph{perturbative} gauge coupling unification, we must restrict the size of the $\chi$ and $\bar\chi$ representations.
Using the parameterization of \eqss{chiN}{chiNbar}{nq}, this requires \cite{Morrissey:2005uz}
\begin{equation} \label{gcc}
N_q \lesssim 6
\end{equation}

\subsection{Baryogenesis}

The potential derived from \eq{W} is a simplified version of \eq{Voriginal}
\begin{eqnarray}
V & = & V_0 - \tilde{m}_\phi^2 |\phi|^2 + m_L^2 |l|^2 - m_{H_u}^2 |h_u|^2 + m_{H_d}^2 |h_d|^2
\nonumber \\ & & {}
+ \left( \frac{1}{2} A_\nu \lambda_\nu l^2 h_u^2 + B \lambda_\mu \phi^2 h_u h_d + \textrm{c.c.} \right)
\nonumber \\ & & {}
+ \left| \lambda_\nu l h_u^2 \right|^2
+ \left| \lambda_\nu l^2 h_u + \lambda_\mu \phi^2 h_d \right|^2
+ \left| \lambda_\mu \phi^2 h_u \right|^2
+ \left| 2 \lambda_\mu \phi h_u h_d \right|^2
\nonumber \\ \label{V} & & {}
+ \frac{1}{2} g^2 \left( |h_u|^2 - |h_d|^2 - |l|^2 \right)^2
\end{eqnarray}
where, as described in Section~\ref{potential},
\begin{equation}
\fn{\tilde{m}_\phi^2}{|\phi|} = \fn{f}{\frac{1}{2} \alpha_\phi \ln\frac{\left|\phi\right|^2 + m_\mathrm{s}^2}{m_\mathrm{s}^2}} m_\phi^2
\end{equation}
and we have set $d=e=0$ since, as discussed in Section~\ref{introbaryo}, they will be held at the origin throughout the dynamics.
Despite setting $\lambda_\phi = 0$, the potential still contains all the terms needed for our baryogenesis scenario, and we expect the thermal inflation and baryogenesis to proceed as described in Section~\ref{introti}.
The detailed differences will be due to the presence of the axion and the suppressed flaton vacuum mass, \eq{flatonmass}, both of which may affect the preheating and decay of the flaton.
The PQ strings formed at the end of thermal inflation may also have some effect, possibly prolonging the leptogenesis.

In Section~\ref{numerical} we describe and present the results of our numerical simulation performed in order to test in detail whether the baryogenesis proceeds as we expect, especially that the preheating leads to a conserved lepton number, as we did for our original model of Ref.~\cite{Jeong:2004hy} in Ref.~\cite{Felder:2007iz}.

\subsection{Flaton decay}

The flaton decays to Standard Model (SM) particles and axions
\begin{equation}
\Gamma_\phi = \Gamma_\mathrm{SM} + \Gamma_a
\end{equation}

\subsubsection{Axion branching ratio}
\label{abr}

\begin{figure}[ht]
\newcommand{\xmin}{-1}
\newcommand{\xmax}{1}
\newcommand{\ymin}{-5}
\newcommand{\ymax}{0}
\begin{tikzpicture}[x=0.4\textwidth,y=0.1067\textwidth]
\loghelplines[black!20]{0,1}{-4,-3,-2,-1,0}
\axes{\displaystyle \frac{m_\mathrm{PQ}}{m_h}}{\Gamma_a/\Gamma_\mathrm{SM}}
\logaxislabels{-1,0,1}{-5,-4,-3,-2,-1,0}
\clip (\xmin,\ymin) rectangle (\xmax,\ymax);
\draw[red,thick] plot file {data/Gamma_a.dat};
\end{tikzpicture}
\caption{ \label{branches}
$\Gamma_a / \Gamma_\mathrm{SM}$ versus $m_\mathrm{PQ} / m_h$ for $m_A = 2|B|$, $|\mu| = 10^3 \GeV$ and $m_h = 125 \GeV$.
}
\end{figure}
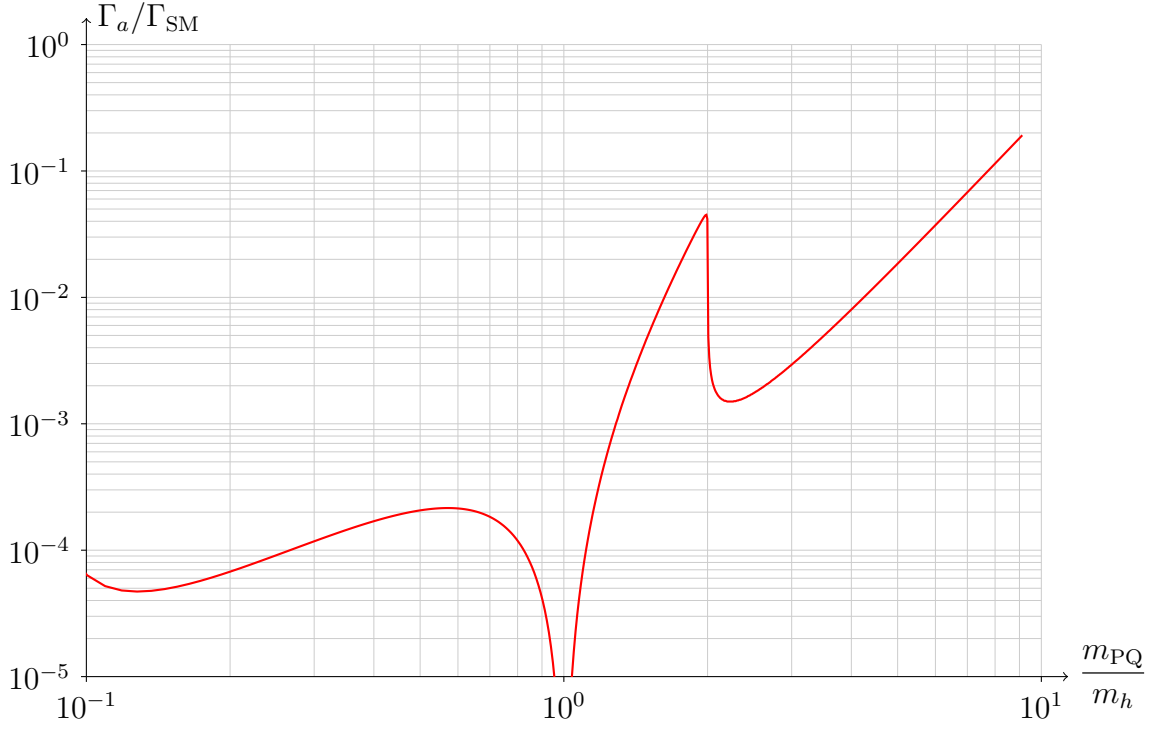

The decay rate to (hot) axions is (see Appendix~\ref{appendix})
\begin{equation} \label{gammaphia}
\Gamma_a = \frac{m_\mathrm{PQ}^3}{64\pi \phi_0^2}
\end{equation}
The decay to SM particles is dominated by three channels: direct decay to SM Higgs, decay to bottom quarks via flaton-Higgs mixing, and decay to gluons via a $\chi \bar\chi$ loop
\begin{equation}
\Gamma_\mathrm{SM} = \Gamma_{\phi \to h h} + \Gamma_{\phi \to b \bar{b}} + \Gamma_{\phi \to g g}
\end{equation}
The decay rates are calculated in Appendix~\ref{appendix}, giving
\begin{eqnarray}
\frac{\Gamma_{\phi \to h h}}{\Gamma_a}
& = & 16 \left( \frac{m_A^2 - |B|^2}{m_A^2} \right)^2 \left( \frac{|\mu|}{m_\mathrm{PQ}} \right)^4 \left[ \sqrt[\mathrm{Re}]{ 1 - \frac{4m_h^2}{m_\mathrm{PQ}^2} } + \order{\gamma_h} + \order{\gamma_h^2} \right]
\\
\frac{\Gamma_{\phi \to b \bar{b}}}{\Gamma_a}
& \sim & \frac{200 m_b^2}{m_h^2} \left( \frac{m_A^2 - |B|^2}{m_A^2} \right)^2 \left( \frac{|\mu|}{m_\mathrm{PQ}} \right)^4 \left( 1 - \frac{4m_b^2}{m_\mathrm{PQ}^2} \right)^\frac{3}{2} \frac{m_h^2}{m_\mathrm{PQ}^2} \left( 1 - \frac{m_h^2}{m_\mathrm{PQ}^2} \right)^{-2}
\\
\frac{\Gamma_{\phi \to g g}}{\Gamma_a}
& \sim & \mathinner{0.1} \alpha_3^2 N_q^2
\end{eqnarray}
where $\gamma_h \sim 10^{-5}$ is defined in \eq{littlegammah}.
For $m_h \sim 125 \GeV$ and $m_\mathrm{PQ}^2 \simeq \alpha_\phi m_\phi^2$, with $\alpha_\phi \sim 0.05$ and $m_\phi \sim |\mu| \sim m_\mathrm{s} \sim 10^2 \textrm{ to } 10^3 \GeV$, we have roughly
\begin{eqnarray}
\frac{\Gamma_{\phi \to h h}}{\Gamma_a}
& \sim & \left\{
\begin{array}{lrc}
10^3 \textrm{ to } 10^4 & \textrm{for} & m_\mathrm{PQ} > 2 m_h
\\
10^{-1} \textrm{ to } 10^{-7} & \textrm{for} & m_\mathrm{PQ} < 2 m_h
\end{array}
\right.
\\
\frac{\Gamma_{\phi \to b \bar{b}}}{\Gamma_a}
& \sim & 10^1 \textrm{ to } 10^2
\\
\frac{\Gamma_{\phi \to g g}}{\Gamma_a}
& \sim & 10^{-2} \textrm{ to } 10^{-3}
\end{eqnarray}
so that either $\phi \to h h$ or $\phi \to b \bar{b}$ dominate the SM channel with $\phi \to g g$ always being negligible.
Thus the SM axion branching ratio has the form
\begin{equation} \label{gsmga}
\frac{\Gamma_\mathrm{SM}}{\Gamma_a} \sim 16 \left( \frac{m_A^2 - |B|^2}{m_A^2} \right)^2 \left( \frac{|\mu|}{m_\mathrm{PQ}} \right)^4 \fn{f}{ \frac{m_h^2}{m_\mathrm{PQ}^2} }
\end{equation}
where
\begin{equation}
\fn{f}{x} = \sqrt[\mathrm{Re}]{1-4x} + \frac{\varepsilon x}{(1-x)^2} \left( 1 - \frac{\varepsilon x}{3} \right)^\frac{3}{2}
\end{equation}
and
\begin{equation}
\varepsilon \sim \frac{12 m_b^2}{m_h^2} \sim 0.02
\end{equation}
see Figure~\ref{branches}.

The branching ratio has a strong and somewhat complicated dependence on $m_h^2 / m_\mathrm{PQ}^2$, with the axion production suppressed if
\footnote{This comes close to providing an anthropic explanation for the little hierarchy, since the Higgs thresholds could lead to sharp jumps in hot axion production, and hence via Big Bang nucleosynthesis in helium abundance, with anthropically relevant effects on stellar evolution.
Unfortunately, the flaton-Higgs mixing channel makes this effect too small.}
\begin{equation}
m_h^2 \lesssim m_\mathrm{PQ}^2 \simeq \alpha_\phi m_\phi^2 \ll m_\mathrm{s}^2
\end{equation}
The overall amplitude of axion production depends on the ratio
\begin{equation}
\left( \frac{m_\mathrm{PQ}}{|\mu|} \right)^4 \simeq \frac{\alpha_\phi^2 m_\phi^4}{|\mu|^4}
\end{equation}
which is highly suppressed in our model, see \eqs{alphatoy}{mphimu}.

\subsubsection{Flaton decay evolution}

While the flaton is decaying, its energy density evolves as \cite{KT} \footnote{We neglect radiation from the AD sector.}
\begin{equation} \label{rhoflaton}
\dot\rho_\phi + 3 H \rho_\phi = - \Gamma_\phi \rho_\phi
\end{equation}
while the entropy increases as
\begin{equation} \label{Sdecay}
\frac{\dot{S}}{S} = \frac{\Gamma_\mathrm{SM} \rho_\phi}{s T}
\end{equation}
with
\begin{equation} \label{Hdecay}
3 H^2 = \rho_\phi + \rho_\mathrm{r}
\end{equation}
and
\begin{equation}
\rho_\mathrm{r} = \rho_\mathrm{SM} + \rho_a
\end{equation}
\eq{rhoflaton} is easily solved to give
\begin{equation} \label{rhophi}
\rho_\phi \propto a^{-3} e^{- \Gamma_\phi t}
\end{equation}
while \eq{Sdecay} gives
\begin{equation}
s^{4/3} = \frac{4}{3} \left( \frac{2\pi^2}{45} \right)^\frac{1}{3} \frac{\Gamma_\mathrm{SM}}{a^4} \int_0^t g_*^{1/3} a^4 \rho_\phi \d{t}
\end{equation}
Treating $g_*$ as slowly varying gives
\begin{equation} \label{rhosmgconst}
\rho_\mathrm{SM} = \frac{\Gamma_\mathrm{SM}}{a^4} \int_0^t a^4 \rho_\phi \d{t}
\end{equation}
and
\begin{equation} \label{sgconst}
s^{4/3} = \frac{4}{3} \left( \frac{2\pi^2 g_*}{45} \right)^\frac{1}{3} \frac{\Gamma_\mathrm{SM}}{a^4} \int_0^t a^4 \rho_\phi \d{t}
\end{equation}
Numerical solutions for $\rho_\phi$, $\rho_\mathrm{r}$ and $\left[ \fn{g_*^{1/4}}{T_\mathrm{d}} / \fn{g_*^{1/4}}{T} \right] \left( S / S_\mathrm{f} \right)$ are shown in Figure~\ref{fig:energydensities}.

\begin{figure}[p]
\centering
\begin{tikzpicture}[x=0.36\textwidth,y=0.08\textwidth]
\newcommand{\xmin}{-1}
\newcommand{\xmax}{1}
\newcommand{\ymin}{-3}
\newcommand{\ymax}{3}
\loghelplines[black!20]{0,1}{-2,-1,0,1,2,3}
\axes{\Gamma_\phi t}{}
\logaxislabels{-1,0,1}{-3,-2,-1,0,1,2,3}
\draw[help lines,red] (0.0257,\ymin) -- (0.0257,\ymax) node[above]{$\Gamma_\phi t_\mathrm{d}$};
\clip (\xmin,\ymin) rectangle (\xmax,\ymax);
\draw[blue,thick] plot file {data/phi.dat};
\node[blue] at (-0.75,2) {flaton};
\draw[violet,thick] plot file {data/radiation.dat};
\draw[violet,thick,dotted] (-1,0.9031) -- (1,-1.0969);
\node[violet] at (0.8,-2.25) {radiation};
\draw[green,thick] plot file {data/S.dat};
\draw[green,thick,dotted] (-1,-1.3557) -- (1,1.1443);
\node[green] at (0.8,-0.3) {entropy};
\end{tikzpicture}
\caption{ \label{fig:energydensities}
Flaton and radiation energy densities and entropy during the flaton decay: $\color{blue} \rho_\phi/\Gamma_\phi^2$, $\color{violet} \rho_\mathrm{r}/\Gamma_\phi^2$, $\color{green} \left[ \fn{g_*^{1/4}}{T_\mathrm{d}} / \fn{g_*^{1/4}}{T} \right] \left( S / S_\mathrm{f} \right)$ versus $\Gamma_\phi t$.
The dotted lines are the early time asymptotic solutions of \eqs{rasympnorm}{Sasympnorm}.
The time corresponding to our definition of $T_\mathrm{d}$ in \eq{rhorTddef} is marked in red.
}
\begin{tikzpicture}[x=0.36\textwidth,y=0.16\textwidth]
\newcommand{\xmin}{-1}
\newcommand{\xmax}{1}
\newcommand{\ymin}{0}
\newcommand{\ymax}{3}
\loghelplines[black!20]{0,1}{1,2,3}
\axes{\displaystyle \frac{m_\mathrm{PQ}}{m_h}}{T_\mathrm{d}}
\logaxislabels{-1,0,1}{0,1,2,3}
\clip (\xmin,\ymin) rectangle (\xmax,\ymax);
\draw[red,thick] plot file {data/T_d.dat};
\end{tikzpicture}
\caption{ \label{fig:Td}
Flaton decay temperature versus flaton mass: $T_\mathrm{d}$ versus $m_\mathrm{PQ} / m_h$ for $\phi_0 = 10^{11} \GeV$, $|\mu| = 10^3 \GeV$, $m_h = 125 \GeV$ and $m_A = 2|B|$.
}
\end{figure}

\paragraph{Flaton decay temperature}

There is no unique decay temperature, instead we define the flaton decay temperature $T_\mathrm{d}$ by
\begin{equation} \label{rhorTddef}
\fn{\rho_\mathrm{r}}{T_\mathrm{d}} \equiv \frac{1}{2} \Gamma_\phi^2
\end{equation}
or
\begin{equation} \label{Tddef}
\frac{\pi^2}{30} \fn{g_*}{T_\mathrm{d}} T_\mathrm{d}^4
= \fn{\rho_\mathrm{SM}}{T_\mathrm{d}}
= \frac{1}{2} \Gamma_\mathrm{SM} \Gamma_\phi
\end{equation}
The time
\begin{equation}
t_\mathrm{d} \simeq \mathinner{1.061} \Gamma_\phi^{-1}
\end{equation}
corresponding to $T_\mathrm{d}$ is marked in Figure~\ref{fig:energydensities}.
At this time
\begin{equation}
\fn{\rho_\phi}{t_\mathrm{d}} \simeq \mathinner{0.495} \Gamma_\phi^2
\end{equation}
and after this time the entropy increases by a factor
\begin{equation} \label{SfSd}
\frac{S_\mathrm{f}}{S_\mathrm{d}} \simeq 1.946
\end{equation}

Substituting the results of Section~\ref{abr} into \eq{Tddef} gives
\begin{eqnarray}
T_\mathrm{d} & \simeq &
\left( \frac{5}{8 \pi^4 \fn{g_*}{T_\mathrm{d}}} \right)^\frac{1}{4}
\left| \frac{m_A^2 - |B|^2}{m_A^2} \right|
\frac{|\mu|^2}{m_\mathrm{PQ}^{1/2} \phi_0}
\left[ \fn{f}{ \frac{m_h^2}{m_\mathrm{PQ}^2} } \right]^\frac{1}{2}
\\ \label{Trsim}
& \simeq & 100 \GeV
\left| \frac{m_A^2 - |B|^2}{m_A^2} \right|
\left( \frac{10^{11} \GeV}{\phi_0} \right)
\left( \frac{|\mu|}{10^3 \GeV} \right)^2
\left( \frac{10^2 \GeV}{m_\mathrm{PQ}} \right)^\frac{1}{2}
\left[ \fn{f}{ \frac{m_h^2}{m_\mathrm{PQ}^2} } \right]^\frac{1}{2}
\nonumber \\
\end{eqnarray}
which is plotted in Figure~\ref{fig:Td}.

\paragraph{Early time decay evolution}
\label{etde}

At early times, $\Gamma_\phi t \ll 1$,
\begin{equation} \label{Hasymp}
3 H^2 \simeq \rho_\phi \propto a^{-3}
\end{equation}
therefore from \eq{rhosmgconst}
\begin{equation} \label{rasymp}
\rho_\mathrm{SM} \simeq \frac{6}{5} \Gamma_\mathrm{SM} H
\propto a^{-3/2}
\end{equation}
and
\begin{equation} \label{Sasymp}
s \simeq \left( \frac{8}{5} \right)^\frac{3}{4} \left( \frac{2\pi^2}{45} \right)^\frac{1}{4} g_*^{1/4} \left( \Gamma_\mathrm{SM} H \right)^{3/4}
\propto a^{-9/8}
\end{equation}
Therefore, using \eq{rhophi},
\begin{eqnarray}
\rho_\phi & \simeq & \left[ \frac{\fn{\rho_\phi}{t_\mathrm{d}} e^{\Gamma_\phi t_\mathrm{d}}}{\Gamma_\phi^2} \right] \Gamma_\phi^2 \left( \frac{a}{a_\mathrm{d}} \right)^{-3}
\simeq \mathinner{1.430} \Gamma_\phi^2 \left( \frac{a}{a_\mathrm{d}} \right)^{-3}
\\ \label{Hasympnorm}
H & \simeq & \left[ \frac{\fn{\rho_\phi}{t_\mathrm{d}} e^{\Gamma_\phi t_\mathrm{d}}}{3\Gamma_\phi^2} \right]^\frac{1}{2} \Gamma_\phi \left( \frac{a}{a_\mathrm{d}} \right)^{-\frac{3}{2}}
\simeq \mathinner{0.690} \Gamma_\phi \left( \frac{a}{a_\mathrm{d}} \right)^{-\frac{3}{2}}
\\ \label{rasympnorm}
\rho_\mathrm{SM} & \simeq & \frac{6}{5} \left[ \frac{\fn{\rho_\phi}{t_\mathrm{d}} e^{\Gamma_\phi t_\mathrm{d}}}{3\Gamma_\phi^2} \right]^\frac{1}{2} \Gamma_\mathrm{SM} \Gamma_\phi \left( \frac{a}{a_\mathrm{d}} \right)^{-\frac{3}{2}}
\simeq \mathinner{0.828} \Gamma_\mathrm{SM} \Gamma_\phi \left( \frac{a}{a_\mathrm{d}} \right)^{-\frac{3}{2}}
\\ \label{Sasympnorm}
\frac{\fn{g_*^{1/4}}{T_\mathrm{d}} S}{\fn{g_*^{1/4}}{T} S_\mathrm{f}} & \simeq & \frac{S_\mathrm{d}}{S_\mathrm{f}} \left( \frac{12}{5} \right)^\frac{3}{4} \left[ \frac{\fn{\rho_\phi}{t_\mathrm{d}} e^{\Gamma_\phi t_\mathrm{d}}}{3\Gamma_\phi^2} \right]^\frac{3}{8} \left( \frac{a}{a_\mathrm{d}} \right)^\frac{15}{8}
\simeq \mathinner{0.750} \left( \frac{a}{a_\mathrm{d}} \right)^\frac{15}{8}
\end{eqnarray}

\section{Constraints}
\label{constraints}

Our model has a rich phenomenology generating many constraints.
In this section we analyze the constraints on our model arising from the MSSM, neutrino masses, thermal inflation, baryogenesis, dark matter and nucleosynthesis.

\subsection{MSSM}

Stability of $LH_u$ in our vacuum requires
\begin{equation} \label{mumlhu}
|\mu|^2 > 2 m_{LH_u}^2
\end{equation}
where $m_{LH_u}^2$ is defined in \eq{key}.
The MSSM $\mu$ parameter is generated by the flaton's vacuum expectation value and has magnitude
\begin{equation} \label{mu}
|\mu| = \left| \lambda_\mu \phi_0^2 \right|
\end{equation}

\subsection{Neutrino masses}

The term $\frac{1}{2} \lambda_\nu (LH_u)^2$ in \eq{W} generates neutrino masses
\begin{equation}
m_\nu = \left| \lambda_\nu H_u^2 \right| = \left| \lambda_\nu \right| v^2 \sin^2{\beta}
\end{equation}
where $v = 174 \GeV$ is the electroweak symmetry breaking scale.
The observed neutrino mixing \cite{Yao:2006px}
\begin{eqnarray}
\Delta m^2_{32} & \simeq & 2.4 \times 10^{-3} \eV^2
\\
\Delta m^2_{21} & \simeq & 8.0 \times 10^{-5} \eV^2
\end{eqnarray}
suggest neutrino masses
\begin{eqnarray}
m_3 & \sim & 5 \times 10^{-2} \eV
\\
m_2 & \sim & 9 \times 10^{-3} \eV
\\
m_1 & \lesssim & m_2
\end{eqnarray}
and so
\begin{equation} \label{lambdanu}
\left| \lambda_\nu \right| \simeq 800 \left( \frac{m_\nu}{10^{-2} \eV} \right) \left( \frac{1}{\sin^2{\beta}} \right)
\end{equation}

\subsection{Thermal inflation and baryogenesis}

For the flaton $\phi$ and AD flat direction $LH_u$ to be unstable during thermal inflation, we require their soft supersymmetry breaking mass squareds to be negative
\begin{equation}
- m_\phi^2 < 0
\end{equation}
and
\begin{equation}
- m^2_{LH_u} < 0
\end{equation}
The critical temperatures of the flaton and AD phase transitions are
\begin{equation}
T_\phi = \frac{m_\phi}{\beta_\phi}
\end{equation}
and
\begin{equation}
T_{LH_u} = \frac{m_{LH_u}}{\beta_{LH_u}}
\end{equation}
where
\begin{equation}
\beta_\phi^2 = \frac{1}{4} \sum_\chi \left| \lambda_\chi \right|^2
\end{equation}
and
\begin{equation}
\beta_{LH_u}^2 = \frac{1}{4} \left( 3 \left| \lambda_t \right|^2 + 3 g_2^2 + g_1^2 \right)
\end{equation}
For the AD phase transition to occur before the flaton phase transition, we require
\begin{equation}
T_\phi < T_{LH_u}
\end{equation}
therefore
\begin{equation} \label{leptoconstraint}
\frac{m_\phi}{\beta_\phi} < \frac{m_{LH_u}}{\beta_{LH_u}}
\end{equation}
Combining this with \eq{mumlhu} gives
\begin{equation} \label{mphimu}
m_\phi^2 < \frac{\beta_\phi^2}{\beta_{LH_u}^2} m_{LH_u}^2 < \frac{\beta_\phi^2}{2 \beta_{LH_u}^2} |\mu|^2
\end{equation}
Now \eqss{flatonmass}{lambdachiest}{alphatoy} give
\begin{equation} \label{mpqmu}
m_\mathrm{PQ} \lesssim \frac{|\mu|}{10}
\end{equation}
and $m_\mathrm{PQ}$ should be fairly close to this bound since we expect $m_\phi \sim m_\mathrm{s} \sim |\mu|$, while requiring $m_\phi \sim m_\mathrm{s} \gtrsim 100 \GeV$ gives
\begin{equation} \label{mpqms}
m_\mathrm{PQ} \gtrsim 20 \GeV
\end{equation}

The roll away of the Affleck-Dine field $LH_u$ may reheat the thermal bath and so extend the thermal inflation.
This would dilute the moduli produced before thermal inflation by an extra factor \cite{Felder:2007iz}
\begin{equation} \label{dilution}
\Delta_\mathrm{AD} \sim \frac{1}{\phi_0}
\end{equation}
where we have taken the equality in Eq.~(22) and hence Eq.~(25) of Ref.~\cite{Felder:2007iz}, and noted that $\Gamma_\mathrm{AD} \ll H_0$ so that the AD energy density is reduced by Hubble expansion rather than decay, and probably has a matter-like equation of state.

After the AD phase transition, the AD field $LH_u$ settles to the minimum
\begin{equation}
|l|^2 = |h_u|^2 = l_0^2
\end{equation}
of its potential
\begin{equation}
V = V_0 + m_L^2 |l|^2 - m_{H_u}^2 |h_u|^2
+ \left( \frac{1}{2} A_\nu \lambda_\nu l^2 h_u^2 + \textrm{c.c.} \right)
+ \left| \lambda_\nu l h_u^2 \right|^2
+ \left| \lambda_\nu l^2 h_u \right|^2
+ \frac{1}{2} g^2 \left( |h_u|^2 - |l|^2 \right)^2
\end{equation}
where \footnote{The $|LH_u|$ dependent renormalisation group running of $m_{LH_u}^2$ may reduce this value.}
\begin{equation}
l_0^2 = \frac{ \sqrt{ 12 m_{LH_u}^2 + \left| A_\nu \right|^2 } + \left| A_\nu \right| }{ 6 \left| \lambda_\nu \right| }
\end{equation}
\eq{lambdanu} gives
\begin{equation}
l_0 \sim 10^9 \GeV \sqrt{ \left( \frac{m_{LH_u}}{10^3 \GeV} \right) \left( \frac{10^{-2} \eV}{m_\nu} \right) }
\end{equation}

For the flaton to be able to bring the AD field $LH_u$ back in to the origin at the end of thermal inflation, we require the flaton potential energy to dominate the AD potential energy
\begin{equation} \label{flatondom}
\alpha_\phi m_\phi^2 \phi_0^2 \gtrsim m^2_{LH_u} l_0^2
\end{equation}
The extra dilution factor of \eq{dilution} is sufficient to effectively remove the lower bound on $\phi_0$ coming from requiring that thermal inflation sufficiently dilute the moduli, compared with the constraint
\begin{equation}
\phi_0 \gg l_0 \sim 10^9 \GeV
\end{equation}
required by \eq{flatondom}.
The upper bound derived from \eq{modulicon} still applies, and using \eq{V0} gives
\begin{equation}
\phi_0 \lesssim 10^{13} \GeV \sqrt{
\left( \frac{n_\mathrm{mod}/s}{10^{-12}} \right)
\left( \frac{m_\mathrm{mod}}{10^2 \GeV} \vphantom{\frac{n_\mathrm{mod}/s}{10^{-12}}} \right)^3
\left( \frac{10^2 \GeV}{m_\mathrm{PQ}} \right)^2
\left( \frac{10 \GeV}{T_\mathrm{d}} \right) }
\end{equation}

The number of $e$-folds of thermal inflation is \cite{Lyth:1995ka}
\begin{equation} \label{efolds}
N \sim \frac{1}{6} \ln \left( \frac{V_0^2}{m_\phi^6 m_\mathrm{s}} \right) + \frac{1}{3} \ln \left( \frac{1}{\phi_0} \right) \sim 11
\end{equation}
where the second term comes from \eq{dilution}.

The baryon asymmetry generated in this model is \cite{Jeong:2004hy,Felder:2007iz}
\begin{eqnarray}
\lefteqn{
\frac{n_B}{s} \sim \left( \frac{n_L}{n_\mathrm{AD}} \right) \left( \frac{m^2_{LH_u} l_0^2}{\alpha_\phi m_\phi^2 \phi_0^2} \right) \left( \frac{T_\mathrm{d}}{m_{LH_u}} \right)
} \\
& \sim & 10^{-10}
\left( \frac{n_L / n_\mathrm{AD}}{10^{-2}} \right)
\left( \frac{10^{12} \GeV}{\phi_0} \right)^2
\left( \frac{T_\mathrm{d}}{1 \GeV} \right)
\left( \frac{10^{-1}}{\alpha_\phi} \right)
\left( \frac{10^{-2} \eV}{m_\nu} \right)
\left( \frac{m_{LH_u}}{m_\phi} \right)^2
\nonumber \\ \label{nBs}
\end{eqnarray}
Comparing with the observed value \cite{Yao:2006px}
\begin{equation}
\frac{n_B}{s} = 9 \times 10^{-11}
\end{equation}
gives the constraint
\begin{equation} \label{baryophicon}
\phi_0 \sim 10^{12} \GeV
\sqrt{
\left( \frac{n_L / n_\mathrm{AD}}{10^{-2}} \right)
\left( \frac{T_\mathrm{d}}{1 \GeV} \right)
\left( \frac{10^{-1}}{\alpha_\phi} \right)
\left( \frac{10^{-2} \eV}{m_\nu} \right)
}
\left( \frac{m_{LH_u}}{m_\phi} \right)
\end{equation}
Our simulation results, Figure~\ref{leptonnumber} and Ref.~\cite{Felder:2007iz}, suggest that $n_L / n_\mathrm{AD} \lesssim 10^{-2}$ and so we get the rough upper bound
\begin{equation} \label{baryophibound}
\phi_0 \lesssim 10^{12} \GeV
\end{equation}
with values near this bound presumably being more natural.

\subsection{Axinos produced by the flaton decay}
\label{flatonaxinos}

The flaton dependent renormalization of the axino mass generates the effective radial flaton axino interaction
\begin{equation} \label{flatonaxinocoupling}
\frac{\alpha_\axino m_\axino}{\sqrt{2}\, \phi_0} \mathinner{\delta r} \axino^2 + \textrm{c.c.}
\end{equation}
where
\begin{equation}
\alpha_\axino \sim 10^{-1}
\end{equation}
is calculated in Appendix~\ref{flatonaxinodecay}.
The flaton decay rate to axinos is
\begin{equation} \label{gammaphiaxino}
\Gamma_{\phi \to \axino\axino} = \frac{\alpha_\axino^2 m_\axino^2 m_\mathrm{PQ}}{32\pi \phi_0^2}
\end{equation}
Using \eq{rhophi}, the flaton decay generates an axino number density
\begin{equation}
n_\axino = \frac{2 \Gamma_{\phi \to \axino\axino}}{m_\mathrm{PQ} a^3} \int_0^t a^3 \rho_\phi \d{t}
= \frac{2 \Gamma_{\phi \to \axino\axino}}{m_\mathrm{PQ} \Gamma_\phi} \left( e^{\Gamma_\phi t} - 1 \right) \rho_\phi
\end{equation}
and so, using Eqs.~(\ref{Tddef}) to (\ref{SfSd}), a late time axino abundance
\begin{equation}
\frac{n_\axino}{s} = \frac{2 \Gamma_{\phi \to \axino\axino}}{m_\mathrm{PQ} \Gamma_\phi} \frac{S_\mathrm{d} e^{\Gamma_\phi t_\mathrm{d}} \fn{\rho_\phi}{t_\mathrm{d}}}{S_\mathrm{f} s_\mathrm{d}}
= \frac{\mathinner{2.2} T_\mathrm{d} \Gamma_{\phi \to \axino\axino}}{m_\mathrm{PQ} \Gamma_\mathrm{SM}}
\end{equation}

\paragraph{Current abundance}

Using \eqs{Tddef}{gammaphiaxino}, the current axino abundance is
\begin{eqnarray}
\Omega_\axino & \simeq & 5.6 \times 10^8 \left( \frac{m_\axino}{1 \GeV} \right) \frac{n_\axino}{s}
\\ \label{flatonaxinocon}
& \simeq & \mathinner{0.36} \frac{\Gamma_\phi^{1/2}}{\Gamma_\mathrm{SM}^{1/2}}
\left( \frac{10}{\fn{g_*^{1/2}}{T_\mathrm{d}}} \right)
\left( \frac{\alpha_\axino}{10^{-1}} \right)^2
\left( \frac{m_\axino}{1 \GeV} \right)^3
\left( \frac{10 \GeV}{T_\mathrm{d}} \right)
\left( \frac{10^{11} \GeV}{\phi_0} \right)^2
\end{eqnarray}
Therefore $\Omega_\axino \leq \Omega_\mathrm{CDM} \simeq 0.2$ requires
\begin{equation}
m_\axino \lesssim 1.8 \GeV \left[
\frac{\Gamma_\mathrm{SM}^{1/2}}{\Gamma_\phi^{1/2}}
\left( \frac{\fn{g_*^{1/2}}{T_\mathrm{d}}}{10} \right)
\left( \frac{T_\mathrm{d}}{1 \GeV} \right)
\left( \frac{10^{-1}}{\alpha_\axino} \vphantom{\frac{T_\mathrm{d}}{1 \GeV}} \right)^2
\left( \frac{\phi_0}{10^{12} \GeV} \right)^2
\right]^\frac{1}{3}
\end{equation}
which is at the lower end of its expected range \eq{axinomass}.

\subsection{Axinos produced by the decay of thermally generated \nobreak{NLSPs}}

\subsubsection{NLSP decay to axinos}

The superpotential coupling $\lambda_\mu \phi^2 H_u H_d$ leads to the decay of NLSPs to axinos via various channels with rate \cite{Covi:1999ty,Covi:2001nw}
\begin{equation} \label{gammahiggsino}
\Gamma_{\nlsp \to \axino} = A \frac{m_\nlsp^3}{16\pi \phi_0^2}
\end{equation}
where we have neglected the decay products' masses, and $A \sim 1$ may contain a factor of $m_Z^2/m_\nlsp^2$.

\subsubsection{Axino abundance}

The thermal bath will generate NLSPs with number density
\begin{eqnarray}
n_\nlsp & = & \frac{1}{\pi^2} \int_0^\infty \frac{ k^2 \d{k} }{ \exp\sqrt{\frac{k^2+m_\nlsp^2}{T^2}} + 1 }
\\
& \equiv & m_\nlsp^3 \fn{f_\nlsp}{\frac{m_\nlsp}{T}}
\end{eqnarray}
whose decays will generate a late time axino abundance
\begin{equation} \label{ndecay}
\frac{n_\axino}{s} = \frac{S_\mathrm{d}}{S_\mathrm{f}} \frac{1}{s_\mathrm{d}} \int_0^\infty \left( \frac{a}{a_\mathrm{d}} \right)^3 n_\nlsp \Gamma_{\nlsp \to \axino} \d{t}
\end{equation}

\paragraph{High $T_\mathrm{d}$ limit}

In the high $T_\mathrm{d}$ limit
\begin{eqnarray}
\frac{n_\axino}{s}
& = & \frac{\Gamma_{\nlsp \to \axino}}{\fn{s}{m_\nlsp}} \int_0^\infty \frac{a^3 n_\nlsp \d{t}}{\fn{a}{m_\nlsp}^3}
\\
& = & \frac{m_\nlsp^3 \Gamma_{\nlsp \to \axino}}{\fn{s}{m_\nlsp} \fn{H}{m_\nlsp}} \int_0^\infty \left( \frac{m_\nlsp}{T} \right)^5 \fn{f_\nlsp}{\frac{m_\nlsp}{T}} \frac{\d{T}}{T}
\\
& = & \left[ \frac{135 \sqrt{5}\,}{\sqrt{2}\, \pi^3} \int_0^\infty x^4 \fn{f_\nlsp}{x} \d{x} \right] \frac{\Gamma_\mathrm{SM}^{1/2} \Gamma_{\nlsp \to \axino}}{\Gamma_\phi^{1/2} \fn{g_*^{3/2}}{m_\nlsp} m_\nlsp^2}
\\ \label{nasanhi}
& = & \frac{5.4}{\fn{g_*^{3/2}}{m_\nlsp}} \frac{\Gamma_\mathrm{SM}^{1/2}}{\Gamma_\phi^{1/2}} \frac{\Gamma_{\nlsp \to \axino}}{m_\nlsp^2}
\end{eqnarray}
where we have used
\begin{equation}
\int_0^\infty x^4 \fn{f_\nlsp}{x} \d{x} = \frac{15 \fn{\zeta}{5}}{2\pi^2} \simeq 0.788
\end{equation}

\paragraph{Low $T_\mathrm{d}$ limit}

Using \eq{rasymp}, the axino production rate
\begin{equation}
- \frac{d (a^3 n_\axino)}{d\ln T}
\propto \frac{a^3 n_\nlsp}{H}
\propto \frac{n_\nlsp}{T^{12}}
\end{equation}
and so, using
\begin{equation}
n_\nlsp \simeq \frac{m_\nlsp^{3/2} T^{3/2}}{\sqrt{2\pi^3}\,} \exp \left( - \frac{m_\nlsp}{T} \right)
\end{equation}
the maximum axino production occurs at
\begin{equation}
T \simeq T_\nlsp = \frac{2}{21} m_\nlsp
\end{equation}
with our low $T_\mathrm{d}$ limit corresponding to $T_\mathrm{d} \ll T_\nlsp$.

We can analytically estimate the axino abundance in \eq{ndecay} using Eqs.~(\ref{Tddef}), (\ref{SfSd}), (\ref{rasymp}) and (\ref{rasympnorm}),
\begin{eqnarray}
\frac{n_\axino}{s}
& = & \frac{8 \fn{S}{T_\mathrm{d}} \fn{a}{T_\nlsp}^3 m_\nlsp^3 \Gamma_{\nlsp \to \axino}}{3 \fn{S}{T_\mathrm{f}} \fn{a}{T_\mathrm{d}}^3 \fn{s}{T_\mathrm{d}} \fn{H}{T_\nlsp}} \int_0^\infty \left( \frac{T_\nlsp}{T} \right)^{12} \fn{f_\nlsp}{\frac{m_\nlsp}{T}} \frac{\d{T}}{T}
\\
& = & \left[ \frac{576 \sqrt{15}\, (0.828)^2}{\pi^3 (1.946)} \int_0^\infty x^{11} \fn{f_\nlsp}{x} \d{x} \right] \frac{\Gamma_\mathrm{SM}^{1/2} \fn{g_*^{3/2}}{T_\mathrm{d}} T_\mathrm{d}^7 \Gamma_{\nlsp \to \axino}}{\Gamma_\phi^{1/2} \fn{g_*^3}{T_\nlsp} m_\nlsp^9}
\\ \label{nasan}
& = & \mathinner{4.4 \times 10^6}
\frac{\fn{g_*^{3/2}}{T_\mathrm{d}}}{\fn{g_*^3}{T_\nlsp}}
\frac{\Gamma_\mathrm{SM}^{1/2}}{\Gamma_\phi^{1/2}}
\frac{\Gamma_{\nlsp \to \axino}}{m_\nlsp^2}
\left( \frac{T_\mathrm{d}}{m_\nlsp} \right)^7
\end{eqnarray}
where we have used
\begin{equation}
\int_0^\infty x^{11} \fn{f_\nlsp}{x} \d{x} = \frac{7}{3} \left( 1 - 2^{-11} \right) \left| B_{12} \right| \pi^{11} \simeq 1.737 \times 10^5
\end{equation}

\paragraph{Numerical}

A numerical solution of \eq{ndecay} can be obtained by writing it in the form
\begin{equation} \label{naxino}
\frac{n_\axino}{s} \simeq \fn{g_*^{-\frac{1}{4}}}{T_\mathrm{d}} \fn{g_*^{-\frac{5}{4}}}{T_\nlsp} \frac{\Gamma_\mathrm{SM}^{1/2}}{\Gamma_\phi^{1/2}} \frac{\Gamma_{\nlsp \to \axino}}{m_\nlsp^2} \fn{F_\axino}{x_\axino}
\end{equation}
where
\begin{equation} \label{Fa}
\fn{F_\axino}{x_\axino} = \fn{g_*^\frac{1}{4}}{T_\mathrm{d}} \fn{g_*^\frac{5}{4}}{T_\nlsp} \frac{\Gamma_\phi^{1/2}}{\Gamma_\mathrm{SM}^{1/2}} \frac{S_\mathrm{d}}{S_\mathrm{f}} \frac{m_\nlsp^2}{s_\mathrm{d}} \int_0^\infty \left( \frac{a}{a_\mathrm{d}} \right)^3 n_\nlsp \d{t}
\end{equation}
versus
\begin{equation} \label{xd}
x_\axino = \frac{2}{3} \frac{\fn{g_*^{1/4}}{T_\mathrm{d}} T_\mathrm{d}}{\fn{g_*^{1/4}}{T_\nlsp} T_\nlsp}
\end{equation}
is plotted in Figure~\ref{fig:axino}.

\begin{figure}[ht]
\centering
\newcommand{\xmin}{-1}
\newcommand{\xmax}{1}
\newcommand{\ymin}{-7}
\newcommand{\ymax}{1}
\begin{tikzpicture}[x=0.4\textwidth,y=0.067\textwidth]
\loghelplines[black!20]{0,1}{-6,-5,-4,-3,-2,-1,0,1}
\axes{x_\axino}{F_\axino}
\logaxislabels{-1,0,1}{-7,-6,-5,-4,-3,-2,-1,0,1}
\clip (\xmin,\ymin) rectangle (\xmax,\ymax);
\draw[red,thick] plot file {data/F_axino.dat};
\draw[red,thick,dotted] (-1,-6.272) -- (1,7.728);
\draw[red,thick,dotted] (-1,0.734) -- (1,0.734);
\end{tikzpicture}
\caption{ \label{fig:axino}
Axino abundance versus flaton decay temperature: $F_\axino$ versus $x_\axino$, see Eqs.~(\ref{naxino}) to (\ref{Faapproxhi}). $F_\axino$, and hence the flaton decay temperature $T_\mathrm{d}$, is constrained by \eq{Traxino}.
}
\end{figure}

For $x_\axino \ll 1$, \eq{nasan} gives
\begin{equation} \label{Faapprox}
\fn{F_\axino}{x_\axino} \sim \mathinner{5.3} x_\axino^7
\end{equation}
while for $x_\axino \gg 1$, \eq{nasanhi} gives
\begin{equation} \label{Faapproxhi}
\fn{F_\axino}{x_\axino} \sim 5.4
\end{equation}

\paragraph{Current abundance}

Using \eqs{gammahiggsino}{naxino}, the current abundance is
\begin{eqnarray}
\Omega_\axino & \simeq & 5.6 \times 10^8 \left( \frac{m_\axino}{1 \GeV} \right) \frac{n_\axino}{s}
\\
& \simeq & \mathinner{270} A
\frac{\Gamma_\mathrm{SM}^{1/2}}{\Gamma_\phi^{1/2}}
\left( \frac{10^3}{\fn{g_*^{1/4}}{T_\mathrm{d}} \fn{g_*^{5/4}}{T_\nlsp}} \right)
\left( \frac{m_\nlsp}{10^2 \GeV} \right)
\left( \frac{m_\axino}{1 \GeV} \right)
\left( \frac{10^{11} \GeV}{\phi_0} \right)^2
\fn{F_\axino}{x_\axino}
\nonumber \\
\end{eqnarray}
Therefore $\Omega_\axino \leq \Omega_\mathrm{CDM} \simeq 0.2$ requires
\begin{equation} \label{Traxino}
\fn{F_\axino}{x_\axino} \lesssim 7.5 \times 10^{-4}
\frac{1}{A}
\frac{\Gamma_\phi^{1/2}}{\Gamma_\mathrm{SM}^{1/2}}
\left( \frac{\fn{g_*^{1/4}}{T_\mathrm{d}} \fn{g_*^{5/4}}{T_\nlsp}}{10^3} \right)
\left( \frac{10^2 \GeV}{m_\nlsp} \right)
\left( \frac{1 \GeV}{m_\axino} \right)
\left( \frac{\phi_0}{10^{11} \GeV} \right)^2
\end{equation}
which can be converted into a constraint on $x_\axino$, and hence the flaton decay temperature $T_\mathrm{d}$, using Figure~\ref{fig:axino}.
For $x_\axino \ll 1$, \eq{Faapprox} gives
\begin{equation} \label{approxomeganlsp}
\Omega_\axino \sim \mathinner{0.19} A
\frac{\Gamma_\mathrm{SM}^{1/2}}{\Gamma_\phi^{1/2}}
\left( \frac{10^3 \fn{g_*^{3/2}}{T_\mathrm{d}}}{\fn{g_*^3}{T_\nlsp}} \right)
\left( \frac{m_\nlsp}{10^2 \GeV} \right)
\left( \frac{m_\axino}{1 \GeV} \right)
\left( \frac{10^{11} \GeV}{\phi_0} \right)^2
\left( \frac{25 T_\mathrm{d}}{m_\nlsp} \right)^7
\end{equation}
and
\begin{equation} \label{anaxioncon}
T_\mathrm{d} \lesssim \frac{m_\nlsp}{25}
\left[ \frac{\fn{g_*}{T_\nlsp}}{\fn{g_*}{T_\mathrm{d}}} \right]^\frac{1}{4}
\left[ \frac{1}{A}
\frac{\Gamma_\phi^{1/2}}{\Gamma_\mathrm{SM}^{1/2}}
\left( \frac{\fn{g_*^{1/4}}{T_\mathrm{d}} \fn{g_*^{5/4}}{T_\nlsp}}{10^3} \right)
\left( \frac{10^2 \GeV}{m_\nlsp} \right)
\left( \frac{1 \GeV}{m_\axino} \right)
\left( \frac{\phi_0}{10^{11} \GeV} \right)^2
\right]^\frac{1}{7}
\end{equation}
Note that the flaton decay temperature will be less than or similar to the neutralino freeze out temperature, which is about $m_\nlsp/20$ \cite{Yao:2006px}.
Therefore we may require
\begin{equation} \label{mpqmn}
m_\mathrm{PQ} < 2 m_\nlsp
\end{equation}
to avoid direct production of neutralinos in the flaton decay.

Axinos will also be produced by the decay of NLSPs after they freeze out.
However, the standard Big Bang neutralino freeze out abundance is good match to the dark matter abundance, our freeze out NLSP abundance will typically be less than the standard abundance, and $m_\axino \ll m_\nlsp$, therefore the axino abundance generated after freeze out should be safe.

\subsection{Cold axions}
\label{axcdm}

\subsubsection{Misalignment axions}

The axion mass turns on around $T \sim \Lambda_\mathrm{QCD}$, see \eq{axionmass}, and the axion starts to oscillate at a temperature $T_\mathrm{o}$ given by \cite{KT}
\begin{equation} \label{osc}
\fn{m_a}{T_\mathrm{o}} \simeq 3 H_\mathrm{o}
\end{equation}
with axion number density
\begin{equation}
\fn{n_a}{T_\mathrm{o}} \propto \fn{m_a}{T_\mathrm{o}}
\end{equation}
and conserved axion number.
The late time misalignment axion abundance is
\begin{equation}
\frac{n_a}{s} = \frac{\fn{S}{T_\mathrm{o}}}{S_\mathrm{f}} \frac{\fn{n_a}{T_\mathrm{o}}}{\fn{s}{T_\mathrm{o}}}
\end{equation}
The abundance is suppressed by a factor $\Delta_a$ compared with the standard Big Bang abundance, where
\begin{equation} \label{deltaadef}
\Delta_a \equiv \frac{(n_a/s)^\mathrm{BB}}{n_a/s}
= \frac{S_\mathrm{f}}{\fn{S}{T_\mathrm{o}}} \frac{\fn{m_a}{T_o^\mathrm{BB}}/\fn{s}{T_o^\mathrm{BB}}}{\fn{m_a}{T_\mathrm{o}}/\fn{s}{T_\mathrm{o}}}
= \frac{\fn{g_*}{T_\mathrm{o}}}{\fn{g_*}{T_\mathrm{o}^\mathrm{BB}}} \left( \frac{T_\mathrm{o}}{T_\mathrm{o}^\mathrm{BB}} \right)^{6.7} \frac{S_\mathrm{f}}{\fn{S}{T_\mathrm{o}}}
\end{equation}
\eqs{axionmass}{osc} give the standard Big Bang axion oscillation temperature
\begin{eqnarray}
T_\mathrm{o}^\mathrm{BB}
& \simeq & \left[ \frac{ 0.1 \fn{m_a}{0} \Lambda_\mathrm{QCD}^{3.7} \left( T_\mathrm{o}^\mathrm{BB} \right)^2 }{ 3 H_\mathrm{o}^\mathrm{BB} } \right]^\frac{1}{5.7}
\\ \label{tobb}
& \simeq & 1.3 \GeV \left( \frac{10^{11}\GeV}{f_a} \right)^\frac{1}{5.7}
\end{eqnarray}

\paragraph{Low $T_\mathrm{d}$ limit}

Our axion oscillation temperature can be analytically estimated using the results of Section~\ref{etde} if the oscillation starts well before the flaton decay completes.
For $T_\mathrm{d} \ll T_\mathrm{o}$, using \eqs{Tddef}{rasymp},
\begin{eqnarray}
T_\mathrm{o}^{7.7} & = & \frac{ H_\mathrm{o}^\mathrm{BB} \left( T_\mathrm{o}^\mathrm{BB} \right)^{3.7} T_\mathrm{o}^4 }{ H_\mathrm{o} }
\\ \label{To}
& = & \frac{6\sqrt{2}\,}{5\sqrt{3}\,} \frac{\Gamma_\mathrm{SM}^{1/2}}{\Gamma_\phi^{1/2}} \frac{\fn{g_*^{1/2}}{T_\mathrm{o}^\mathrm{BB}} \fn{g_*^{1/2}}{T_\mathrm{d}}}{\fn{g_*}{T_\mathrm{o}}} \left( T_\mathrm{o}^\mathrm{BB} \right)^{5.7} T_\mathrm{d}^2
\end{eqnarray}
and, using \eqs{rasympnorm}{Sasympnorm},
\begin{equation} \label{SfSo}
\frac{S_\mathrm{f}}{\fn{S}{T_\mathrm{o}}} = \mathinner{0.71} \left[ \frac{\fn{g_*}{T_\mathrm{d}}}{\fn{g_*}{T_\mathrm{o}}} \right]^\frac{1}{4} \left[ \frac{\fn{\rho_\mathrm{SM}}{T_\mathrm{o}}}{\fn{\rho_\mathrm{SM}}{T_\mathrm{d}}} \right]^\frac{5}{4}
= \mathinner{0.71} \frac{\fn{g_*}{T_\mathrm{o}}}{\fn{g_*}{T_\mathrm{d}}} \left( \frac{T_\mathrm{o}}{T_\mathrm{d}} \right)^5
\end{equation}
Therefore, for $T_\mathrm{d} \ll T_\mathrm{o}$, the misalignment axion abundance is suppressed by a factor
\begin{equation} \label{deltaa}
\Delta_a = \mathinner{0.71} \left( \frac{6\sqrt{2}\,}{5\sqrt{3}\,} \right)^\frac{11.7}{7.7} \left( \frac{\Gamma_\mathrm{SM}}{\Gamma_\phi} \right)^\frac{11.7}{2 \times 7.7} \left[ \frac{ \fn{g_*^2}{T_\mathrm{o}} }{ \fn{g_*}{T_\mathrm{o}^\mathrm{BB}} \fn{g_*}{T_\mathrm{d}} } \right]^\frac{3.7}{2 \times 7.7}
\left( \frac{T_\mathrm{o}^\mathrm{BB}}{T_\mathrm{d}} \right)^\frac{3 \times 4.7 + 1}{7.7}
\end{equation}

\paragraph{High $T_\mathrm{d}$ limit}

In the opposite limit, $T_\mathrm{d} \gg T_\mathrm{o}$, dilution due to the flaton decay is negligible but the presence of hot axions produced in the flaton decay increases $H$ for a given $T$, and so reduces $T_\mathrm{o}$, slightly enhancing the cold axion production.
For $T_\mathrm{d} \gg T_\mathrm{o}$
\begin{equation} \label{Tohigh}
T_\mathrm{o} = \left[ \frac{ \fn{g_*}{T_\mathrm{o}^\mathrm{BB}} }{ \fn{g_*}{T_\mathrm{o}} } \frac{\Gamma_\mathrm{SM}}{\Gamma_\phi} \right]^\frac{1}{2 \times 5.7} T_\mathrm{o}^\mathrm{BB}
\end{equation}
and
\begin{equation} \label{Deltaahigh}
\Delta_a = \left[ \frac{ \fn{g_*}{T_\mathrm{o}} }{ \fn{g_*}{T_\mathrm{o}^\mathrm{BB}} } \right]^\frac{4.7}{2 \times 5.7} \left( \frac{\Gamma_\mathrm{SM}}{\Gamma_\phi} \right)^\frac{6.7}{2 \times 5.7}
\end{equation}

\paragraph{Numerical}

We can numerically solve \eq{osc} using \eqss{rhoflaton}{Sdecay}{Hdecay} to give
\begin{equation}
\left( \frac{\Gamma_\phi}{\Gamma_\mathrm{SM}} \right)^\frac{1}{2 \times 5.7} \left[ \frac{\fn{g_*}{T_\mathrm{o}^\mathrm{BB}}}{\fn{g_*}{T_\mathrm{o}}} \right]^\frac{3.7}{4 \times 5.7} \frac{\fn{g_*^{1/4}}{T_\mathrm{o}} T_\mathrm{o}}{\fn{g_*^{1/4}}{T_\mathrm{o}^\mathrm{BB}} T_\mathrm{o}^\mathrm{BB}}
= \fn{F_\mathrm{o}}{x_a}
\end{equation}
where
\begin{equation} \label{xo}
x_a = \left( \frac{\Gamma_\phi}{\Gamma_\mathrm{SM}} \right)^\frac{1}{2 \times 5.7} \left[ \frac{\fn{g_*}{T_\mathrm{o}^\mathrm{BB}}}{\fn{g_*}{T_\mathrm{o}}} \right]^\frac{3.7}{4 \times 5.7} \frac{\fn{g_*^{1/4}}{T_\mathrm{d}} T_\mathrm{d}}{\fn{g_*^{1/4}}{T_\mathrm{o}^\mathrm{BB}} T_\mathrm{o}^\mathrm{BB}}
\end{equation}
For $x_a \ll 1$, \eq{To} gives
\begin{equation}
\fn{F_\mathrm{o}}{x_a} \sim \mathinner{1.0} x_a^{0.26}
\end{equation}
while for $x_a \gg 1$, \eq{Tohigh} gives
\begin{equation}
\fn{F_\mathrm{o}}{x_a} \sim 1
\end{equation}
\eq{deltaadef} then gives the suppression factor
\begin{equation} \label{deltaanum}
\Delta_a = \left( \frac{\Gamma_\mathrm{SM}}{\Gamma_\phi} \right)^\frac{6.7}{2 \times 5.7} \left[ \frac{ \fn{g_*}{T_\mathrm{o}} }{ \fn{g_*}{T_\mathrm{o}^\mathrm{BB}} } \right]^\frac{4.7}{2 \times 5.7} \frac{\fn{g_*^{1/4}}{T_\mathrm{f}}}{\fn{g_*^{1/4}}{T_\mathrm{o}}} \fn{F_{\Delta_a}}{x_a}
\end{equation}
where $T_\mathrm{f} = \min(T_\mathrm{d},T_\mathrm{o})$ and
\begin{equation} \label{FDelta}
\fn{F_{\Delta_a}}{x_a} = \left( \frac{\Gamma_\phi}{\Gamma_\mathrm{SM}} \right)^\frac{6.7}{2 \times 5.7} \left[ \frac{\fn{g_*}{T_\mathrm{o}^\mathrm{BB}}}{\fn{g_*}{T_\mathrm{o}}} \right]^\frac{3.7 \times 6.7}{4 \times 5.7} \left[ \frac{\fn{g_*^{1/4}}{T_\mathrm{o}} T_\mathrm{o}}{\fn{g_*^{1/4}}{T_\mathrm{o}^\mathrm{BB}} T_\mathrm{o}^\mathrm{BB}} \right]^{6.7} \frac{\fn{g_*^{1/4}}{T_\mathrm{o}} S_\mathrm{f}}{\fn{g_*^{1/4}}{T_\mathrm{f}} \fn{S}{T_\mathrm{o}}}
\end{equation}
is plotted in Figure~\ref{fig:FDelta}.

\begin{figure}[ht]
\centering
\newcommand{\xmin}{-1}
\newcommand{\xmax}{1}
\newcommand{\ymin}{-1}
\newcommand{\ymax}{2}
\begin{tikzpicture}[x=0.4\textwidth,y=0.178\textwidth]
\loghelplines[black!20]{0,1}{0,1,2}
\axes{x_a}{F_{\Delta_a}}
\logaxislabels{-1,0,1}{-1,0,1,2}
\clip (\xmin,\ymin) rectangle (\xmax,\ymax);
\draw[red,thick] plot file {data/F_axion.dat};
\draw[red,thick,dotted] (-1,1.7988) -- (1,-2.1212);
\end{tikzpicture}
\caption{ \label{fig:FDelta}
Axion dilution factor versus flaton decay temperature: $F_{\Delta_a}$ versus $x_a$, see \eqss{xo}{deltaanum}{FDelta}.
}
\end{figure}

For $x_a \ll 1$, \eq{deltaa} gives
\begin{equation}
\fn{F_{\Delta_a}}{x_a} \sim \mathinner{0.69} x_a^{-1.96}
\end{equation}
while for $x_a \gg 1$, \eq{Tohigh} gives
\begin{equation}
\fn{F_{\Delta_a}}{x_a} \sim 1
\end{equation}

\subsubsection{String axions}

The axion abundance produced by the decay of PQ strings is \cite{KT}
\begin{equation}
n_a \propto \frac{1}{a^3} \int_0^{a_\mathrm{o}} \frac{a^3 H^2}{\omega} \frac{\d{a}}{a}
\end{equation}
with the typical axion energy $\omega \propto H$.
Therefore, using \eq{osc}, the late time abundance is
\begin{equation}
\frac{n_a}{s} \propto \frac{\fn{S}{T_\mathrm{o}}}{S_\mathrm{f}} \frac{\fn{m_a}{T_\mathrm{o}}}{\fn{s}{T_\mathrm{o}}} \int_0^{a_\mathrm{o}} \frac{a^3 H}{a_\mathrm{o}^3 H_\mathrm{o}} \frac{da}{a}
\end{equation}
Comparing with \eq{deltaadef}, our string axion abundance is suppressed compared with the standard Big Bang abundance by a factor
\begin{equation} \label{deltaastring}
\Delta_a^\mathrm{string} = \left( \int_0^{a_\mathrm{o}} \frac{a^3 H}{a_\mathrm{o}^3 H_\mathrm{o}} \frac{da}{a} \right)^{-1} \Delta_a^\mathrm{misalign}
\end{equation}
For $T_\mathrm{d} \ll T_\mathrm{o}$, \eq{Hasymp} gives
\begin{equation}
\Delta_a^\mathrm{string} \sim \frac{3}{2} \Delta_a^\mathrm{misalign}
\end{equation}
while for $T_\mathrm{d} \gg T_\mathrm{o}$
\begin{equation}
\Delta_a^\mathrm{string} \sim \Delta_a^\mathrm{misalign}
\end{equation}

\subsubsection{Current abundance}

Taking into account the dilution factor $\Delta_a$, \eqs{fama}{fasa} become
\begin{equation} \label{omegaa}
\Omega_a \sim \frac{0.2}{\Delta_a} \left( \frac{f_a}{10^{10} \textrm{ to } 10^{11} \GeV} \right)^{1.175}
\end{equation}
For $T_\mathrm{d} \gg T_\mathrm{o}^\mathrm{BB} \sim 1 \GeV$, the dilution factor $\Delta_a$ is slightly less than one, see \eq{Deltaahigh}, giving a slight enhancement of the cold axion abundance over the standard Big Bang case, which may provide an observational signal of our model in the future.
For $T_\mathrm{d} \ll T_\mathrm{o}^\mathrm{BB}$, \eq{deltaa} gives the dilution factor $\Delta_a \sim (T_\mathrm{d}/T_\mathrm{o}^\mathrm{BB})^{-1.96}$, and so, using \eqss{faphi0}{Trsim}{tobb}, we have $\Omega_a \propto f_a^{-0.44}$.

\subsection{Hot axions}
\label{hotaxion}

Hot axions will be produced in the flaton decay with energy density
\begin{equation} \label{rhoa}
\frac{\rho_a^\mathrm{hot}}{\rho_\mathrm{SM}} = \left[ \frac{\fn{g_*}{T_\mathrm{d}} \fn{g_{*S}^{4/3}}{T}}{\fn{g_*}{T} \fn{g_{*S}^{4/3}}{T_\mathrm{d}}} \right] \frac{\Gamma_a}{\Gamma_\mathrm{SM}}
\end{equation}
where $\Gamma_a/\Gamma_\mathrm{SM}$ is given by \eq{gsmga} and plotted in Figure~\ref{branches}.
The axion production is highly suppressed in our model due to the suppressed flaton mass, but will typically be higher than the thermally produced axion energy density \cite{KT}
\begin{equation}
\frac{\rho_a^\mathrm{th}}{\rho_\mathrm{SM}} \sim 10^{-6} \left( \frac{10^{11} \GeV}{f_a} \right)^2
\end{equation}
The hot axions produced in the flaton decay have current momentum
\begin{equation} \label{pa}
p_a = \frac{a}{a_0} \frac{m_\mathrm{PQ}}{2}
\end{equation}
where $a$ is the scale factor at the time they were created and $a_0$ is the scale factor now.
In particular, the current momentum of an axion produced at $t_\mathrm{d}$ is
\begin{eqnarray}
p_\mathrm{d} & = & \frac{a_\mathrm{d}}{a_0} \frac{m_\mathrm{PQ}}{2}
\\
& = & \left[ \frac{S_\mathrm{d}^{1/3} \fn{g_{*S}^{1/3}}{T_0} T_0}{S_\mathrm{f}^{1/3} \fn{g_{*S}^{1/3}}{T_\mathrm{d}} T_\mathrm{d}} \right] \frac{m_\mathrm{PQ}}{2}
\\
& \simeq & 1.48 \times 10^{-4} \eV \left[ \frac{m_\mathrm{PQ}}{\fn{g_*^{1/3}}{T_\mathrm{d}} T_\mathrm{d}} \right]
\end{eqnarray}
so, comparing with \eq{ma0}, they are marginally relativistic now.
The current number density spectrum is
\begin{equation}
p_a \frac{dn_a^\mathrm{hot}}{dp_a} = \left( \frac{a}{a_0} \right)^3 \frac{2 \rho_\phi}{m_\mathrm{PQ}} \frac{\Gamma_a}{H} = \frac{16 \Gamma_a \Gamma_\phi p_\mathrm{d}^3}{m_\mathrm{PQ}^4} \frac{\rho_\phi p_a^3}{H \Gamma_\phi p_\mathrm{d}^3}
\end{equation}
which may provide an observational test of our model in the future.
Assuming that the hot axions are still relativistic now, their current energy density is
\begin{equation}
\Omega_a^\mathrm{hot} \simeq 2 \times 10^{-5} \left[ \frac{100}{\fn{g_*}{T_\mathrm{d}}} \right]^\frac{1}{3} \frac{\Gamma_a}{\Gamma_\mathrm{SM}}
\end{equation}

\subsection{Nucleosynthesis}
\label{sec:bbn}

The flaton decay should complete before Big Bang nucleosynthesis, requiring
\begin{equation}
T_\mathrm{d} \gtrsim 10 \MeV
\end{equation}

The good agreement between the Standard Model prediction for the ${}^4\mathrm{He}$ mass fraction \cite{Cyburt:2003fe}
\begin{equation}
Y_p^\mathrm{SM} = 0.248
\end{equation}
and the observed value \cite{Yao:2006px}
\begin{equation} \label{Ypobs}
Y_p^\mathrm{obs} = 0.25 \pm 0.01
\end{equation}
constrains any additional contribution to the energy density at the time of nucleosynthesis from hot axions produced in the flaton decay \cite{Cyburt:2004yc}
\begin{equation} \label{BBNcon}
\left. \frac{\rho_a}{\rho_\mathrm{SM}} \right|_\mathrm{BBN} \lesssim 0.14
\end{equation}
\eq{rhoa} then gives the constraint
\begin{equation} \label{abranchbbn}
\frac{\Gamma_a}{\Gamma_\mathrm{SM}} \lesssim 0.3
\end{equation}
which is automatically satisfied in our model, see \eqs{gsmga}{mpqmu}, and Figure~\ref{fig:con1}.

\subsection{Summary of constraints}
\label{consum}

\begin{figure}[p]
\centering
\newcommand{\xmin}{1}
\newcommand{\xmax}{3}
\newcommand{\ymin}{2}
\newcommand{\ymax}{4}
\begin{tikzpicture}[x=0.375\textwidth,y=0.25\textwidth]
\loghelplines{2,3}{3,4}
\begin{scope} \clip (\xmin,\ymin) rectangle (\xmax,\ymax);
\fill[blue,semitransparent] (\xmin,\ymin) -- plot file {data/flatonaxino_fa=11_ma=1.dat} -- (\xmax,\ymin) -- cycle;
\fill[violet,semitransparent] (\xmin,\ymax) -- plot file {data/NLSPaxino_fa=11_mNLSP=200.dat} -- (\xmax,\ymax) -- cycle;
\draw[green!50,thick] plot file {data/axion_fa=11_string.dat};
\draw[blue!50,thick] plot file {data/flatonaxino_fa=11_ma=2.dat};
\draw[violet!50,thick] plot file {data/NLSPaxino_fa=11_mNLSP=100.dat};
\fill[orange!50] (1,2) -- (3,4) -- (3,2) -- cycle;
\fill[red!50] (\xmin,\ymin) -- plot file {data/BBN.dat} -- (\xmax,\ymin) -- cycle;
\fill[yellow!50] (1,2) -- (1,4) -- (1.3010,4) -- (1.3010,2) -- cycle;
\end{scope}
\node[blue] at (1.5,2.74) {axino};
\node[violet] at (2,3.54) {axino};
\node[orange] at (1.9,2.5) {baryogenesis};
\node[yellow text] at (1.15,3) {susy};
\node[red] at (2.6,2.25) {BBN};
\axes{\displaystyle \frac{m_\mathrm{PQ}}{\GeV}}{\mu/\GeV}
\logaxislabels{1,2,3}{2,3,4}
\end{tikzpicture}
\begin{tikzpicture}[x=0.375\textwidth,y=0.25\textwidth]
\loghelplines{2,3}{3,4}
\begin{scope} \clip (\xmin,\ymin) rectangle (\xmax,\ymax);
\fill[green,semitransparent] (\xmin,\ymax) -- plot file {data/axion_fa=12_mis.dat} -- (\xmax,\ymax) -- cycle;
\fill[blue,semitransparent] (\xmin,\ymin) -- plot file {data/flatonaxino_fa=12_ma=1.dat} -- (\xmax,\ymin) -- cycle;
\fill[violet!50] (\xmin,\ymax) -- plot file {data/NLSPaxino_fa=12_mNLSP=200.dat} -- (\xmax,\ymax) -- cycle;
\draw[green!50,thick] plot file {data/axion_fa=12_string.dat};
\draw[blue!50,thick] plot file {data/flatonaxino_fa=12_ma=2.dat};
\draw[violet!50,thick] plot file {data/NLSPaxino_fa=12_mNLSP=100.dat};
\fill[orange!50] (1,2) -- (3,4) -- (3,2) -- cycle;
\fill[red!50] (\xmin,\ymin) -- plot file {data/BBN.dat} -- (\xmax,\ymin) -- cycle;
\fill[yellow!50] (1,2) -- (1,4) -- (1.3010,4) -- (1.3010,2) -- cycle;
\end{scope}
\node[green] at (1.75,3.39) {axion};
\node[violet] at (2.08,3.85) {axino};
\node[orange] at (1.9,2.5) {baryogenesis};
\node[yellow text] at (1.15,3) {susy};
\node[red] at (2.6,2.25) {BBN};
\axes{\displaystyle \frac{m_\mathrm{PQ}}{\GeV}}{\mu/\GeV}
\logaxislabels{1,2,3}{2,3,4}
\end{tikzpicture}
\caption{ \label{fig:con1}
Parameter space constraints, $|\mu|$ versus $m_\mathrm{PQ}$, for $\phi_0 = 3 f_a / \sqrt{2}$, $m_h = 125 \GeV$ and $m_A = 2|B|$.
Top $f_a = 10^{11} \GeV$ and bottom $f_a = 10^{12} \GeV$.
{\color{red!50} $\blacksquare$} hot axion over-abundance at Big Bang nucleosynthesis, \eq{abranchbbn}.
{\color{orange!50} $\blacksquare$} thermal inflation baryogenesis consistency constraint, \eq{mpqmu}.
{\color{yellow!50} $\blacksquare$} flaton mass constraint, \eq{mpqms}.
{\color{green!50} $\blacksquare$} axion cold dark matter over-abundance, \eq{omegaa} for {\color{green!50} solid} misalignment or {\color{green!50} line} high string estimate.
{\color{blue!50} $\blacksquare$}{\color{violet!50} $\blacksquare$} axino cold dark matter over-abundance: {\color{blue!50} $\blacksquare$} \eq{flatonaxinocon} for $\alpha_\axino = 0.1$ and {\color{blue!50} solid} $m_\axino = 1 \GeV$ or {\color{blue!50} line} $m_\axino = 2 \GeV$; {\color{violet!50} $\blacksquare$} \eqs{Traxino}{mpqmn} for $m_\axino = 1 \GeV$ and {\color{violet!50} solid} $m_\nlsp = 200 \GeV$ or {\color{violet!50} line} $m_\nlsp = 100 \GeV$.
}
\end{figure}

\begin{figure}[p]
\centering
\newcommand{\xmin}{2}
\newcommand{\xmax}{4}
\newcommand{\ymin}{9}
\newcommand{\ymax}{13}
\begin{tikzpicture}[x=0.27\textwidth,y=0.09\textwidth]
\loghelplines{3,4}{10,11,12,13}
\begin{scope} \clip (\xmin,\ymin) rectangle (\xmax,\ymax);
\fill[green,semitransparent] (\xmax,\ymax) -- plot file {data/axion_mPQ=0p1_mis.dat} -- cycle;
\fill[blue,semitransparent] (\xmin,\ymin) -- plot file {data/flatonaxino_mPQ=0p1_ma=1.dat} -- (\xmax,\ymin) -- cycle;
\fill[violet!50] (\xmin,\ymin) -- plot file {data/NLSPaxino_mPQ=0p1_mNLSP=200.dat} -- (\xmax,\ymax) -- (\xmax,\ymin) -- cycle;
\fill[violet!50] (3.6021,\ymin) -- (3.6021,\ymax) -- (\xmax,\ymax) -- (\xmax,\ymin) -- cycle;
\draw[green!50,thick] plot file {data/axion_mPQ=0p1_mis.dat};
\draw[green!50,thick] plot file {data/axion_mPQ=0p1_string.dat};
\draw[blue!50,thick] plot file {data/flatonaxino_mPQ=0p1_ma=1.dat};
\draw[blue!50,thick] plot file {data/flatonaxino_mPQ=0p1_ma=2.dat};
\draw[violet!50,thick] plot file {data/NLSPaxino_mPQ=0p1_mNLSP=100.dat};
\fill[orange!50] (2,13) -- (2,12) -- (4,12) -- (4,13) -- cycle;
\fill[yellow!50] (2,9) -- (2,13) -- (2.3010,13) -- (2.3010,9) -- cycle;
\end{scope}
\node[green] at (2.85,11.42) {axion};
\node[blue] at (2.55,10.75) {axino};
\node[violet] at (2.75,9.5) {axino};
\node[orange] at (3.15,12.5) {baryogenesis};
\node[yellow text] at (2.15,11) {susy};
\axes{\displaystyle \frac{\mu}{\GeV}}{f_a/\GeV}
\logaxislabels{2,3,4}{9,10,11,12,13}
\end{tikzpicture}
\begin{tikzpicture}[x=0.27\textwidth,y=0.09\textwidth]
\loghelplines{3,4}{10,11,12,13}
\begin{scope} \clip (\xmin,\ymin) rectangle (\xmax,\ymax);
\fill[green,semitransparent] (\xmax,\ymax) -- plot file {data/axion_mPQ=0p05_mis.dat} -- cycle;
\fill[blue,semitransparent] (\xmin,\ymin) -- plot file {data/flatonaxino_mPQ=0p05_ma=1.dat} -- (\xmax,\ymin) -- cycle;
\fill[violet!50] (\xmin,\ymin) -- plot file {data/NLSPaxino_mPQ=0p05_mNLSP=200.dat} -- (\xmax,\ymax) -- (\xmax,\ymin) -- cycle;
\fill[violet!50] (3.9031,\ymin) -- (3.9031,\ymax) -- (\xmax,\ymax) -- (\xmax,\ymin) -- cycle;
\draw[green!50,thick] plot file {data/axion_mPQ=0p05_mis.dat};
\draw[green!50,thick] plot file {data/axion_mPQ=0p05_string.dat};
\draw[blue!50,thick] plot file {data/flatonaxino_mPQ=0p05_ma=1.dat};
\draw[blue!50,thick] plot file {data/flatonaxino_mPQ=0p05_ma=2.dat};
\draw[violet!50,thick] plot file {data/NLSPaxino_mPQ=0p05_mNLSP=100.dat};
\fill[orange!50] (2,13) -- (2,12) -- (4,12) -- (4,13) -- cycle;
\fill[yellow!50] (2,9) -- (2,13) -- (2.6021,13) -- (2.6021,9) -- cycle;
\end{scope}
\node[green] at (3,11.42) {axion};
\node[blue] at (2.775,10.75) {axino};
\node[violet] at (3,9.5) {axino};
\node[orange] at (3.3,12.5) {baryogenesis};
\node[yellow text] at (2.3,11) {susy};
\axes{\displaystyle \frac{\mu}{\GeV}}{f_a/\GeV}
\logaxislabels{2,3,4}{9,10,11,12,13}
\end{tikzpicture}
\begin{tikzpicture}[x=0.275\textwidth,y=0.09167\textwidth]
\loghelplines{3,4}{10,11,12,13}
\begin{scope} \clip (\xmin,\ymin) rectangle (\xmax,\ymax);
\fill[green,semitransparent] (\xmax,\ymax) -- plot file {data/axion_mPQ=0p025_mis.dat} -- cycle;
\fill[blue,semitransparent] (\xmin,\ymin) -- plot file {data/flatonaxino_mPQ=0p025_ma=1.dat} -- (\xmax,\ymin) -- cycle;
\fill[violet!50] (\xmin,\ymin) -- plot file {data/NLSPaxino_mPQ=0p025_mNLSP=200.dat} -- (\xmax,\ymax) -- (\xmax,\ymin) -- cycle;
\draw[green!50,thick] plot file {data/axion_mPQ=0p025_mis.dat};
\draw[green!50,thick] plot file {data/axion_mPQ=0p025_string.dat};
\draw[blue!50,thick] plot file {data/flatonaxino_mPQ=0p025_ma=1.dat};
\draw[blue!50,thick] plot file {data/flatonaxino_mPQ=0p025_ma=2.dat};
\draw[violet!50,thick] plot file {data/NLSPaxino_mPQ=0p025_mNLSP=100.dat};
\fill[orange!50] (2,13) -- (2,12) -- (4,12) -- (4,13) -- cycle;
\fill[yellow!50] (2,9) -- (2,13) -- (2.9031,13) -- (2.9031,9) -- cycle;
\end{scope}
\node[green] at (3.15,11.42) {axion};
\node[violet] at (3.2,9.5) {axino};
\node[orange] at (3.45,12.5) {baryogenesis};
\node[yellow text] at (2.45,11) {susy};
\axes{\displaystyle \frac{\mu}{\GeV}}{f_a/\GeV}
\logaxislabels{2,3,4}{9,10,11,12,13}
\end{tikzpicture}
\caption{ \label{fig:con2}
Parameter space constraints, $f_a$ versus $|\mu|$.
Top $m_\mathrm{PQ} = \mathinner{0.1} |\mu|$, middle $m_\mathrm{PQ} = \mathinner{0.05} |\mu|$ and bottom $m_\mathrm{PQ} = \mathinner{0.025} |\mu|$.
{\color{orange!50} $\blacksquare$} thermal inflation baryogenesis abundance constraint, \eq{baryophibound}.
}
\end{figure}

Figure~\ref{fig:con1} shows the constraints on $|\mu|$ and $m_\mathrm{PQ}$ for $f_a = 10^{11}$ and $10^{12} \GeV$, while Figure~\ref{fig:con2} shows the constraints on $f_a$ and $|\mu|$ for $m_\mathrm{PQ} = 0.1$, $0.05$ and $0.025 |\mu|$.
There are many strong constraints but nevertheless interesting regions of parameter space survive.
The axino mass is required to be at the lower end of its expected range, $m_\axino \sim 1 \GeV$.
Avoiding over-production of axinos requires $f_a \gtrsim 10^{11} \GeV$, which nicely coincides with the more natural range of parameters for our baryogenesis scenario near the rough upper bound $f_a \lesssim 10^{12} \GeV$.
Avoiding over-production of axions splits the allowed parameter space into two regions.
For $f_a \sim 10^{11} \GeV$, the axion abundance is in good agreement with the observed dark matter abundance, but to avoid over-production of axinos the MSSM $\mu$ parameter is confined to the range $700 \GeV \lesssim |\mu| \lesssim 2 \TeV$.
For $f_a \sim 10^{12} \GeV$, the axino constraint weakens, but to avoid over-production of axions the decay temperature has to be low enough to dilute them, which requires lower values of $|\mu|$ in the range $400 \GeV \lesssim |\mu| \lesssim 800 \GeV$.

The tightness of the dark matter over-production constraints implies that the dark matter should be composed of significant amounts of both axions and axinos.
For example, \eqss{flatonaxinocon}{approxomeganlsp}{omegaa} give
\begin{eqnarray}
\Omega_\axino^\phi & \propto & \phi_0^{-2} T_\mathrm{d}^{-1}
\\
\Omega_\axino^\nlsp & \propto & \phi_0^{-2} T_\mathrm{d}^7
\qquad \left( T_\mathrm{d} \ll \frac{m_\nlsp}{7} \right)
\\
\Omega_a & \propto & \phi_0^{1.175}
\qquad \left( T_\mathrm{d} \gg 1 \GeV \right)
\end{eqnarray}
therefore the dark matter minimum at $f_a \sim 10^{11} \GeV$, $|\mu| \sim 1 \TeV$ has
\begin{equation}
\Omega_\axino^\phi = \mathinner{7} \Omega_\axino^\nlsp
\end{equation}
\begin{equation}
\Omega_a = \mathinner{1.7} \Omega_\axino
\end{equation}
and
\begin{eqnarray}
\lefteqn{
\Omega_{a+\axino} \sim \mathinner{0.2} \times A^{0.05} 
\left( \frac{6}{N} \right)^{0.74}
\left( \frac{\alpha_\axino}{10^{-1}} \right)^{0.65}
\left[ \frac{10^2}{\fn{g_*}{T_\mathrm{d}}} \right]^{0.09}
\left[ \frac{10^2}{\fn{g_*}{T_\nlsp}} \right]^{0.14}
\left( \frac{\Gamma_\phi}{\Gamma_\mathrm{SM}} \right)^{0.14}
} \nonumber \\ && \times
\left( \frac{m_\axino}{1\GeV} \right)^{1.02}
\left( \frac{200\GeV}{m_\nlsp} \right)^{0.28}
\left( \frac{\Lambda_\mathrm{QCD}}{200\MeV} \right)^{0.63}
\left( \frac{s_0/3H_0^2}{5.6 \times 10^8 \GeV^{-1}} \right)
\end{eqnarray}
Alternatively, \eqs{nBs}{omegaa} give
\begin{eqnarray}
\Omega_\mathrm{b} & \propto & T_\mathrm{d} \phi_0^{-2}
\\
\Omega_a & \propto & \phi_0^{1.52} T_\mathrm{d}^{1.96}
\qquad \left( T_\mathrm{d} \ll 1 \GeV \right)
\end{eqnarray}
therefore, fixing $\Omega_\mathrm{b}$, the dark matter minimum at $f_a \sim 10^{12} \GeV$, $|\mu| \sim 600 \GeV$ has
\begin{equation}
\Omega_\axino \simeq \Omega_\axino^\phi = \mathinner{1.4} \Omega_a
\end{equation}
and
\begin{eqnarray}
\lefteqn{
\Omega_{a+\axino} \sim \mathinner{0.2} \times
\left( \frac{6}{N} \right)^{0.64}
\left( \frac{\alpha_\axino}{10^{-1}} \right)^{1.15}
\left[ \frac{10^2}{\fn{g_*}{T_\mathrm{d}}} \right]^{0.29}
\left( \frac{\Gamma_\phi}{\Gamma_\mathrm{SM}} \right)^{0.61}
} \nonumber \\ && \times
\left[
\left( \frac{10^{-2}}{n_L/n_\mathrm{AD}} \right)
\left( \frac{\alpha_\phi}{10^{-1}} \right)
\left( \frac{m_\phi}{m_{LH_u}} \right)^2
\left( \frac{m_\nu}{10^{-2}\eV} \right)
\left( \frac{174\GeV}{v_\mathrm{EW}} \right)^2
\left( \frac{1\GeV}{m_p} \right)
\left( \frac{\Omega_\mathrm{b}}{0.05} \right)
\right]^{0.25} 
\nonumber \\ && \times
\left( \frac{m_\axino}{1\GeV} \right)^{1.73}
\left( \frac{200\MeV}{\Lambda_\mathrm{QCD}} \right)^{0.41}
\left( \frac{s_0/3H_0^2}{5.6 \times 10^8 \GeV^{-1}} \right)^{0.75}
\end{eqnarray}

To need to be at a minimum of the dark matter abundance may seem like tuning, but, taking into account anthropic selection effects, it can be quite natural for our local part of the universe to be found at such extrema of parameter space.
Similarly, our mixed (supersymmetric!) dark matter may also seem odd, but is due to our extremum of parameter space, and follows the observed trend of an over-complicated composition of our local part of the universe, which is again well motivated anthropically.

\section{Numerical simulation of the leptogenesis}
\label{numerical}

We closely follow the methods of Ref.~\cite{Felder:2007iz}.
We performed a three dimensional lattice simulation with periodic boundary conditions using the algorithm described in Ref.~\cite{Felder:2007iz}.
The potential was that of \eq{V} with $f$ given by \eq{f}, the $D$-term
\begin{equation}
D = |h_u|^2 - |h_d|^2 - |l|^2 = \epsilon^2
\end{equation}
and gauge
\begin{equation}
J_\mathbf{a} = \sum_{\psi \in \left\{ h_u , h_d , l \right\} } \frac{ \psi^* Q D_\mathbf{a} \psi - \left( D_\mathbf{a} \psi \right)^* Q \psi }{ i } = 0
\end{equation}
constraints exactly conserved by the algorithm, and canonical gauge invariant kinetic terms.
The numerical parameter $\epsilon$ was introduced to cut off the singularity at $h_u = 0$.
The physical parameters and fields were rescaled by a typical soft supersymmetry breaking mass $m$ and a typical flaton expectation value $M_\mathrm{TI}$ or AD ($l$, $h_u$ or $h_d$) expectation value $M_\mathrm{AD}$, as described in Refs.~\cite{Jeong:2004hy,Felder:2007iz}. As the initial condition, we used the $\Delta = 4 m$ case from Ref.~\cite{Felder:2007iz}.

\subsection{Simulation parameters}

The lattice volume $L^3$, number of lattice points $N^3$, time step $\Delta t$, and $D$-term constraint singularity cutoff $\epsilon$, were taken as \footnote{It was necessary to take $L$ larger than in Ref.~\cite{Felder:2007iz} due to the suppressed flaton mass of \eq{flatonmass}.}
\begin{equation} \label{latticevars}
L = \mathinner{200} m^{-1}
\quad , \quad
N = 128
\quad , \quad
\Delta t = \mathinner{4 \times 10^{-3}} m^{-1}
\quad , \quad
\epsilon = \mathinner{5 \times 10^{-3}} M_\mathrm{AD}
\end{equation}
We tested our results using different values of these numerical parameters.
Limited computing power constrained us to $N \leq 128$.

The $k$-modes allowed by this lattice are $k = \sqrt{k_1^2 + k_2^2 + k_3^2}$ with
\begin{equation}
k_i = \frac{2\pi n_i}{L}
\end{equation}
where $n_i = 0,1,\ldots,N/2$ with $n_i = 0,N/2$ having degeneracy one and the rest degeneracy two.
This spans the range
\begin{equation}
0.033 \, m = \frac{2\pi}{L} \leq k \leq \frac{\sqrt{3}\,\pi N}{L} = \mathinner{3.5} m
\end{equation}

The physical parameters were taken as follows \footnote{The mass squareds were set to three decimal places to avoid any accidental resonances and also to fit the MSSM constraints described in Ref.~\cite{Jeong:2004hy}.}
\begin{equation}
m_{\phi}^2 = \mathinner{0.217} m^2
\quad , \quad
\alpha_\phi = 0.051758
\quad , \quad
m_\mathrm{s}^2 = m_{\phi}^2
\end{equation}
\begin{equation}
m_{H_u}^2 = \mathinner{1.510} m^2
\quad , \quad
m_{H_d}^2 = \mathinner{3.533} m^2
\quad , \quad
m_{L}^2 = \mathinner{1.323} m^2
\end{equation}
\begin{equation}
\left| A_\nu \right| = \mathinner{0.25} m
\quad , \quad
\left| B \right| = \mathinner{1.15} m
\end{equation}
\begin{equation}
\left| \lambda_\nu \right| = \mathinner{0.2} m M_\mathrm{AD}^{-2}
\quad , \quad
\left| \lambda_\mu \right| = \mathinner{2.0} m M_\mathrm{TI}^{-2}
\end{equation}
\begin{equation}
M_\mathrm{AD} = 10^{-2} M_\mathrm{TI}
\end{equation}
\begin{equation} \label{phaseredef}
\arg \left( A_\nu \lambda_\nu \right) = \arg \left( B \lambda_\mu \right) = 0
\end{equation}
Note that the phases of Eq. (\ref{phaseredef}) are not physical and can be adjusted by field rotations.
The CP phase was taken as
\begin{equation} \label{CP}
\arg \left( - B^* A_\nu \right) = \left\{
\begin{array}{ll}
\displaystyle \pi - \frac{\pi}{20} & \qquad CP+ \\
\displaystyle \pi & \qquad CP0 \\
\displaystyle \pi + \frac{\pi}{20} & \qquad CP-
\end{array} \right.
\end{equation}

For the choice of parameters above, we have \cite{Felder:2007iz}
\begin{equation}
\phi_0 = \mathinner{0.70} M_\mathrm{TI}
\quad , \quad
\phi_\mathrm{c} = \mathinner{0.67} \phi_0
\quad , \quad
l_0 = \mathinner{1.5 \times 10^{-2}} \phi_0
\end{equation}
where $|\phi| = \phi_\mathrm{c}$ is the critical value at which the minimum of the AD sector switches to the origin
\begin{equation}
\left( \frac{\phi_\mathrm{c}}{\phi_0} \right)^4 = \frac{m^2_{H_u} - m^2_L}{|\mu|^2}
\end{equation}
The flaton mass squared eigenvalues at $\phi = \phi_0$ are
\begin{equation}
m_\mathrm{PQ}^2 = \alpha_\phi m_\phi^2 = \mathinner{0.011} m^2
\quad , \quad
m^2_a = 0
\end{equation}
and the Affleck-Dine sector mass squared eigenvalues at $\phi = \phi_0$, $l = h_u = h_d = 0$ are
\begin{equation}
m^2_{L H_u} = \mathinner{0.37} m^2
\quad , \quad
m_{H_u H_d}^{2-} = \mathinner{0.54} m^2
\quad , \quad
m_{H_u H_d}^{2+} = \mathinner{3.4} m^2
\end{equation}
and
\begin{equation}
|\mu|^2 = \mathinner{0.94} m^2
\end{equation}
Characteristic potential values are
\begin{equation}
V_0 = \frac{1}{2} \alpha_\phi m_\phi^2 \phi_0^2
\quad , \quad
V_1 = \mathinner{5.9 \times 10^{-3}} V_0
\quad , \quad
V_2 = \mathinner{3.2 \times 10^{-2}} V_0
\end{equation}
where
\begin{equation}
V_1 \equiv V_0 - \fn{V}{0,h_{u0},0,l_0}
\end{equation}
is the depth of the Affleck-Dine sector minimum when the flaton is still at the origin, and
\begin{equation}
V_2 \equiv \fn{V}{\phi_0,h_{u0},0,l_0}
\end{equation}
is the height to which that point is lifted when the flaton reaches its minimum.
Here $\fn{V}{\phi,h_u,h_d,l}$ is the potential.

\subsection{Results}

\begin{figure}[t]
\centering
\includegraphics[width=\textwidth,bb=91 3 322 146]{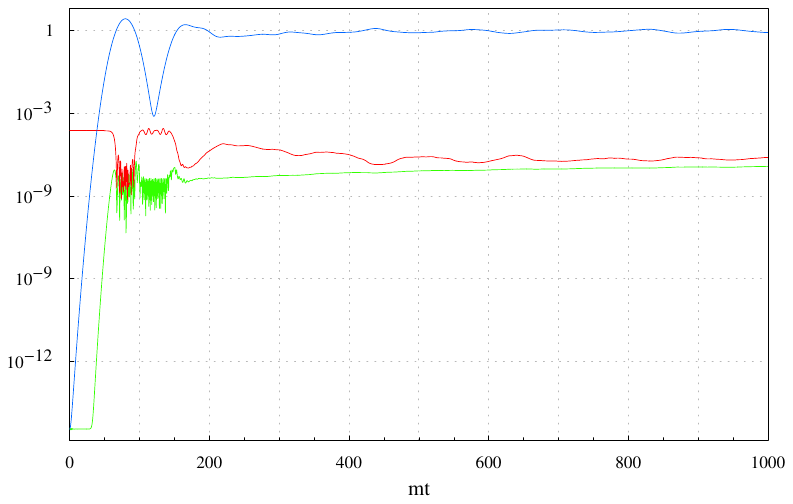}
\caption{ \label{logmag}
Magnitude squareds of the flaton and AD fields averaged over the lattice as a function of time ($CP-$):
$\color{blue} \langle |\phi|^2 \rangle / \phi_0^2$, $\color{red} \langle |l|^2 \rangle / \phi_0^2$, $\color{green} \langle |h_d|^2 \rangle / \phi_0^2$ versus $mt$.
}
\end{figure}

The overall dynamics can be seen in Figure~\ref{logmag}.
Initially, the flaton $\phi$ and the Higgs field $h_d$ are near the origin, the AD field $l$ is near its thermal inflation minimum at $|l|=l_0$, and $h_u$ is determined by $l$ and $h_d$ via the $D$-term constraint.
As $\phi$ rolls away from the origin, ending thermal inflation, it first forces $h_d$ to become non-zero, and then, as it crosses $|\phi| = \phi_\mathrm{c}$, it causes the AD potential to flip, forcing $l$ and $h_d$ in towards the origin.
The AD fields then oscillate about the origin while the flaton rolls beyond its minimum and then back in towards the origin.
Note that the flaton evolves slowly relative to the rapid AD field oscillations since the flaton mass scale $m_\mathrm{PQ} \simeq \alpha_\phi^{1/2} m_\phi$ is suppressed relative to the AD sector mass scales.
As the flaton returns past $|\phi| = \phi_\mathrm{c}$, the AD potential flips back to its original form, forcing $l$ back out to $|l| \sim l_0$ where it oscillates rapidly.
The flaton nears the origin and then rolls out again, crossing $|\phi| = \phi_\mathrm{c}$ and flipping the AD potential for the final time, bringing the AD fields back in again towards the origin.
The flaton then settles to oscillate about its minimum since the build up of gradient energy (preheating) has by this time drained enough kinetic energy from the homogeneous mode to prevent it from returning to $|\phi| < \phi_\mathrm{c}$.
The AD field $l$ initially oscillates around the origin with large amplitude but gradually settles down closer to the origin.

\begin{figure}[t]
\centering
\includegraphics[width=\textwidth,bb=91 3 322 146]{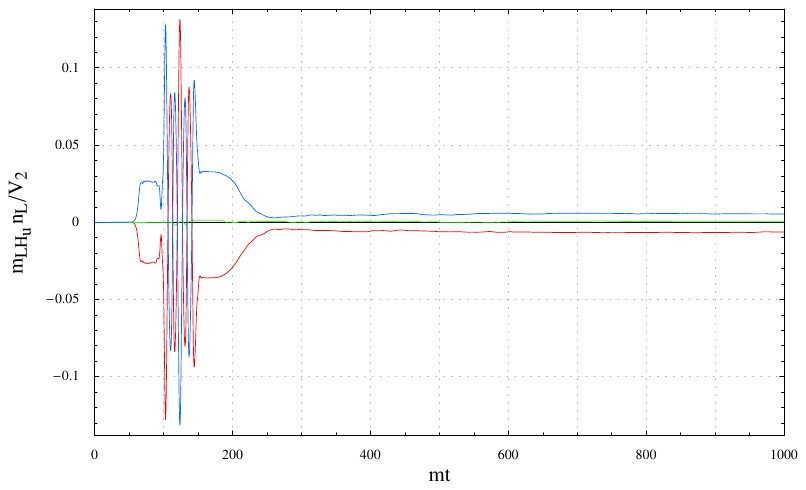}
\caption{ \label{leptonnumber}
Lepton number density averaged over the lattice as a function of time: $m_{LH_u} \langle n_L \rangle / V_2$ ($\color{red} CP+$, $\color{green} CP0$, $\color{blue} CP-$) versus $mt$.
}
\end{figure}

The important outcome of this dynamics is shown in Figure~\ref{leptonnumber}.
The AD field $l$ is initially sitting in its thermal inflation minimum, and hence the lepton number is zero.
The flaton then brings it in towards the origin with rotation, as described in Section~\ref{introbaryo}, and hence lepton number is generated.
This is initially conserved while $l$ is held in the symmetrical lepton number conserving part of its potential near the origin.
Then the flaton forces $l$ back out towards its lepton number violating thermal inflation minimum, about which it oscillates, and hence the lepton number oscillates violently.
The flaton then again brings $l$ back in towards the origin with rotation, and the corresponding lepton number is initially conserved as $l$ is held near the origin by the overextended flaton field.
As the flaton settles down to oscillate about its minimum, the $l$ field is held less tightly and spreads out to feel the lepton violating outer parts of its potential, and hence the lepton number decays.
This decay is halted as $l$ settles down closer to the origin, leaving a residual conserved lepton number.
Although the dynamics is very complicated, Figure~\ref{leptonnumber} shows that the lepton number is controlled by the $CP$ violating phase in our potential, \eq{CP}.

\begin{figure}[p]
\centering
\includegraphics[width=0.95\textwidth,bb=91 3 322 146]{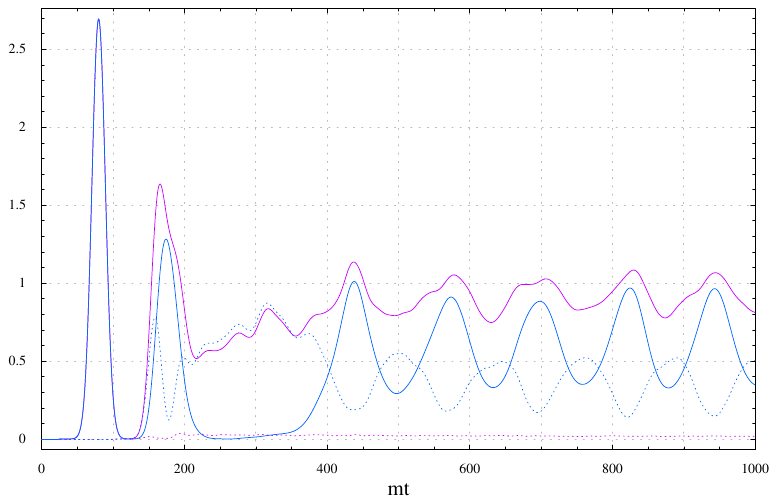}
\caption{ \label{flatonmeanvar}
Mean squareds (solid) and variances (dotted) of the {\color{blue} complex} and {\color{violet} radial} flaton fields as a function of time ($CP-$): $| \langle \phi \rangle |^2 / \phi_0^2$, $\langle | \phi - \langle \phi \rangle |^2 \rangle / \phi_0^2$, $\langle |\phi| \rangle^2 / \phi_0^2$, $\langle ( |\phi| - \langle |\phi| \rangle )^2 \rangle / \phi_0^2$ versus $mt$.
}
\end{figure}

\begin{figure}[p]
\centering
\includegraphics[width=0.95\textwidth,bb=91 3 322 146]{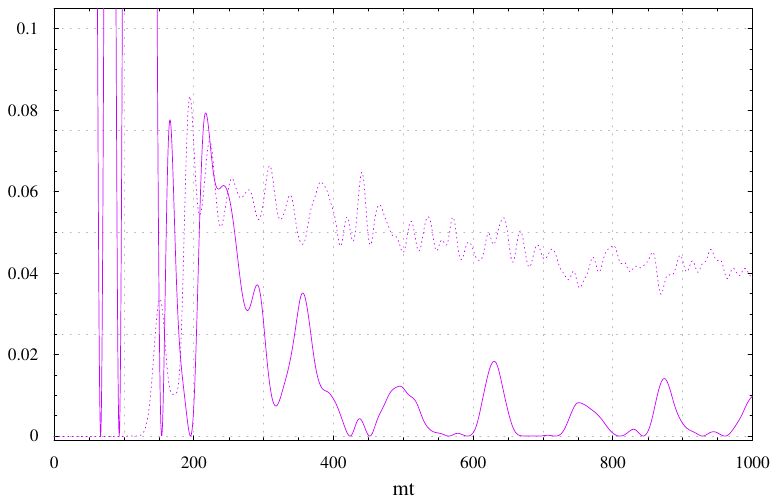}
\caption{ \label{flatonmeanvarmin}
Mean displacement from minimum squared (solid) and variance (dotted) of the radial flaton field as a function of time ($CP-$): $\langle |\phi| - \phi_0 \rangle^2 / \phi_0^2$, $\langle ( |\phi| - \langle |\phi| \rangle )^2 \rangle / \phi_0^2$ versus $mt$.
}
\end{figure}

The flaton dynamics is shown in more detail in Figs.~\ref{flatonmeanvar} and \ref{flatonmeanvarmin}.
The first oscillation of the flaton is essentially homogeneous, but the angular dispersion becomes significant in the second oscillation, limiting its amplitude.
The radial flaton then settles down to oscillate about its minimum, with fairly small amplitude and dispersion, as can be seen in Figure~\ref{flatonmeanvarmin}.
The axion on the other hand settles down to large amplitude oscillations in its dispersion, as can be seen from the  late time oscillations in the mean squared and variance of the complex flaton field in Figure~\ref{flatonmeanvar}.

\begin{figure}[t]
\centering
\includegraphics[width=\textwidth,height=0.6\textheight,bb=91 3 322 204]{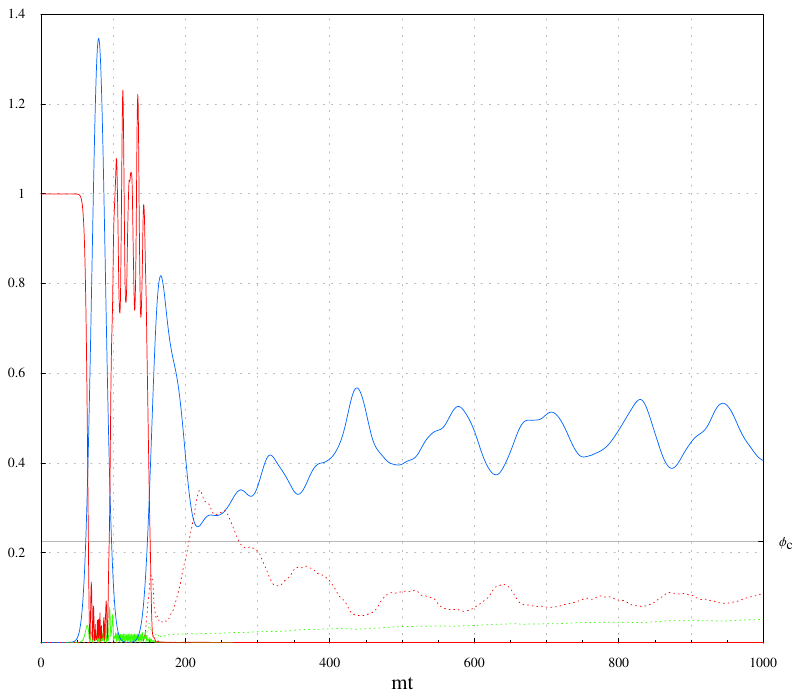}
\caption{ \label{flatonADmeanvar}
Magnitude squared of the flaton field and mean squareds (solid) and variances (dotted) of the AD fields as a function of time ($CP-$): $\color{blue} \langle |\phi|^2 \rangle / 2 \phi_0^2$, $\color{red} | \langle l \rangle |^2 / l_0^2$, $\color{red} \langle | l - \langle l \rangle |^2 \rangle / l_0^2$, $\color{green} | \langle h_d \rangle |^2 / l_0^2$, $\color{green} \langle | h_d - \langle h_d \rangle |^2 \rangle / l_0^2$ versus $mt$.
}
\end{figure}

The effect of the flaton dynamics on the AD sector is shown in more detail in Figure~\ref{flatonADmeanvar}.
The first two large amplitude flaton oscillations force the AD sector in towards the origin, then out, then back in again.
As the flaton settles down, the AD fields mean squareds quickly settle to the origin but the variance now becomes significant.
The variance of $l$ is initially large but importantly decreases over time leading to the conservation of the lepton number seen in Figure~\ref{leptonnumber}.
The variance of $h_d$ on the other hand is initially small but grows over time.

\begin{figure}[t]
\centering
\includegraphics[width=0.49\textwidth,bb=91 3 322 146]{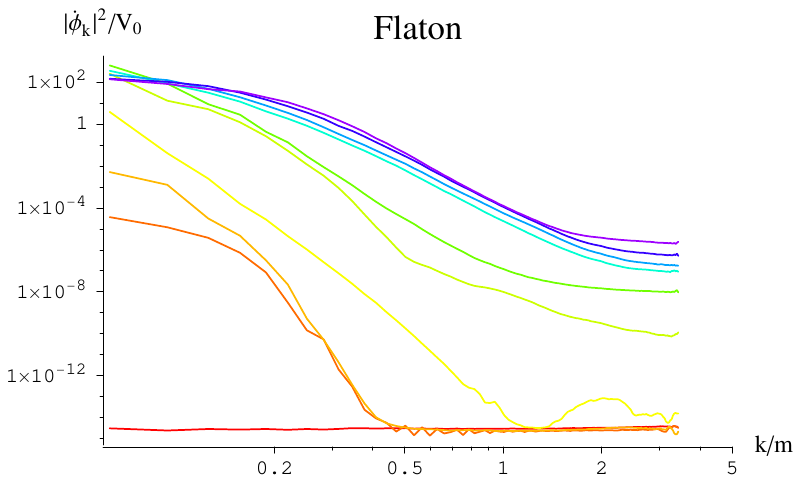}
\includegraphics[width=0.49\textwidth,bb=91 3 322 146]{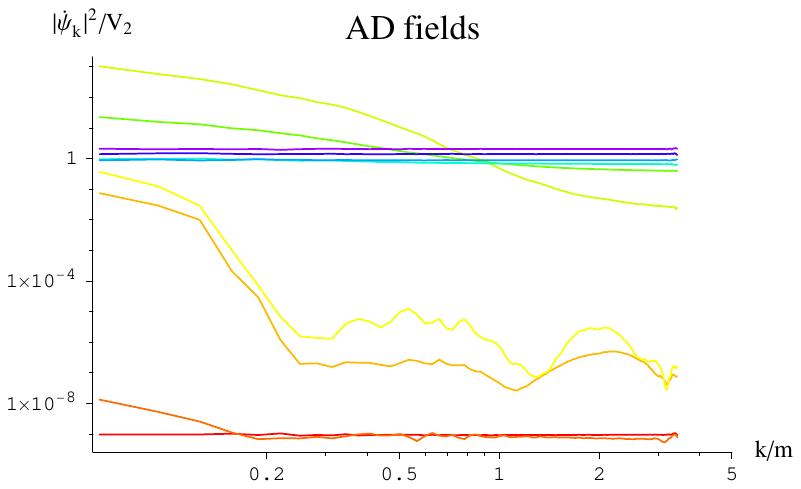}
\includegraphics[width=0.49\textwidth,bb=91 3 322 146]{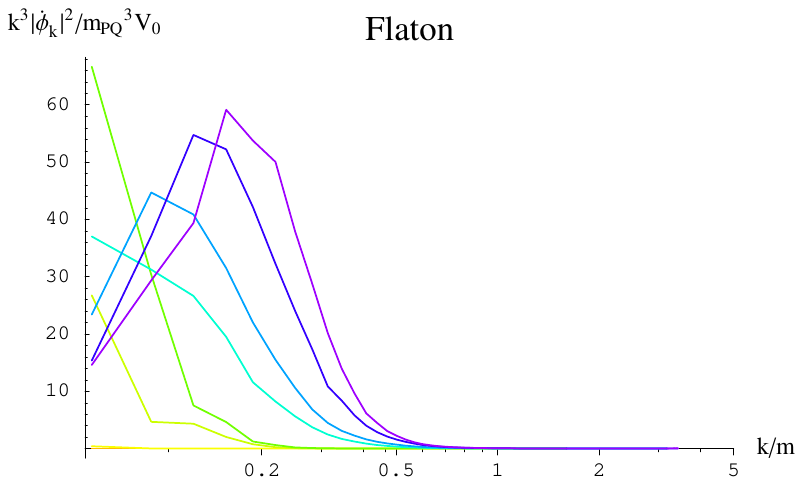}
\includegraphics[width=0.49\textwidth,bb=91 3 322 146]{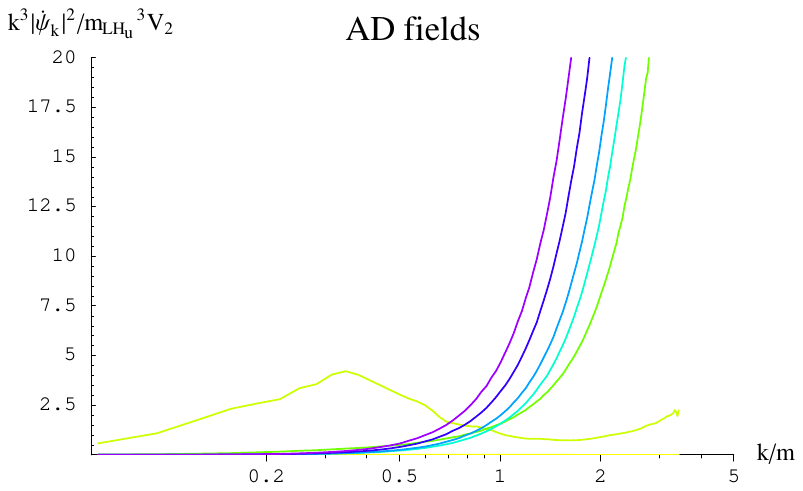}
\caption{ \label{spectra}
Kinetic energy spectra of flaton (left) and AD (right) sectors, thermally normalized (top) so that a thermal spectrum is flat, and energy normalized (bottom) so that the area under the graph is the total energy ($CP-$): $| \dot\phi_k |^2 / V_0$, $| \dot\psi_k |^2 / V_2$, $k^3 | \dot\phi_k |^2 / m_\mathrm{PQ}^3 V_0$, $k^3 | \dot\psi_k |^2 / m_{LH_u}^3 V_2$ at $t = {\color[wave]{681} 0}, {\color[wave]{645} 40}, {\color[wave]{611} 80}, {\color[wave]{578} 120}, {\color[wave]{547} 160}, {\color[wave]{518} 200}, {\color[wave]{491} 250}, {\color[wave]{465} 300}, {\color[wave]{440} 500}, {\color[wave]{417} 1000}$ versus $k/m$.
Relevant mass scales are $m_\mathrm{PQ} = 0.11 \, m$, $m_\phi = 0.47 \, m$, $m_{LH_u} = 0.61 \, m$, $m_{H_uH_d}^- = 0.73 \, m$, $m_{H_uH_d}^+ = 1.8 \, m$.
}
\end{figure}

The preheating of the flaton and AD sectors is shown in Figure~\ref{spectra}.
The initial roll out of the flaton gives rise to tachyonic and angular preheating, exciting modes with $k \lesssim m_\phi$, as can be seen in the top left graph.
Thereafter, the flaton preheating becomes broader, but is still mostly limited to modes with $k \lesssim m_\phi$, and the spectrum settles down to a fairly stable distribution well within the cutoff of our lattice, as can be seen in the bottom left graph.
The AD preheating is more complex, as discussed in Ref.~\cite{Felder:2007iz}, and very efficient.
As can be seen from the top right graph, all the AD lattice modes have thermalized by about $t \sim 250 \, m^{-1}$.
Thus the AD preheating is artificially truncated by our lattice cutoff, so in reality should be even more efficient, bringing the AD fields in even closer to the origin.
However, the preheating available with our limited lattice is sufficient to lead to excellent conservation of lepton number, as shown in Figure~\ref{leptonnumber}.

\begin{figure}[p]
\centering
\includegraphics[width=0.95\textwidth,bb=91 3 322 146]{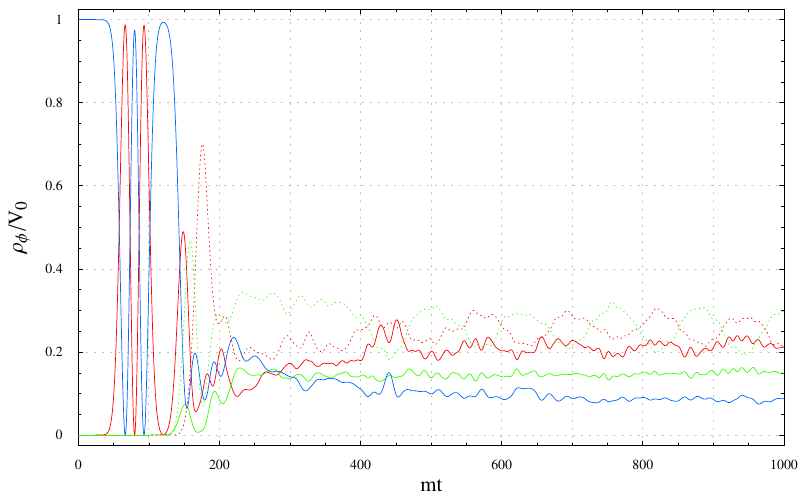}
\caption{ \label{flatonKGPenergy}
{\color{red} Kinetic}, {\color{green} gradient} and {\color{blue} potential} energy densities of the radial (solid) and angular (dotted) flaton fields averaged over the lattice as a function of time ($CP-$): $\rho_\phi / V_0$ versus $mt$.
}
\end{figure}

The energy densities in the flaton sector are shown in Figure~\ref{flatonKGPenergy}.
The first flaton oscillation is dominated by radial kinetic and potential energy, as would be expected for an essentially homogeneous oscillation.
During the second oscillation, a large part of the potential energy is converted into angular gradient and then angular kinetic energy, causing the radial flaton to settle down around its minimum.
The late time axion dispersion oscillations are again evident from the angular kinetic and gradient energies.

\begin{figure}[p]
\centering
\includegraphics[width=0.95\textwidth,bb=91 3 322 146]{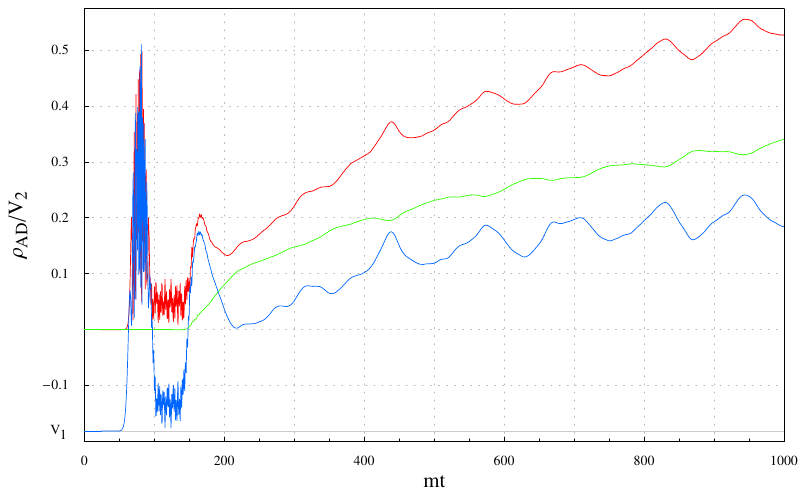}
\caption{ \label{ADKGPenergy}
{\color{red} Kinetic}, {\color{green} gradient} and {\color{blue} potential} energy densities of the AD sector averaged over the lattice as a function of time ($CP-$): $\rho_\mathrm{AD} / V_2$ versus $mt$.
}
\end{figure}

\begin{figure}[p]
\centering
\includegraphics[width=0.95\textwidth,bb=91 3 322 146]{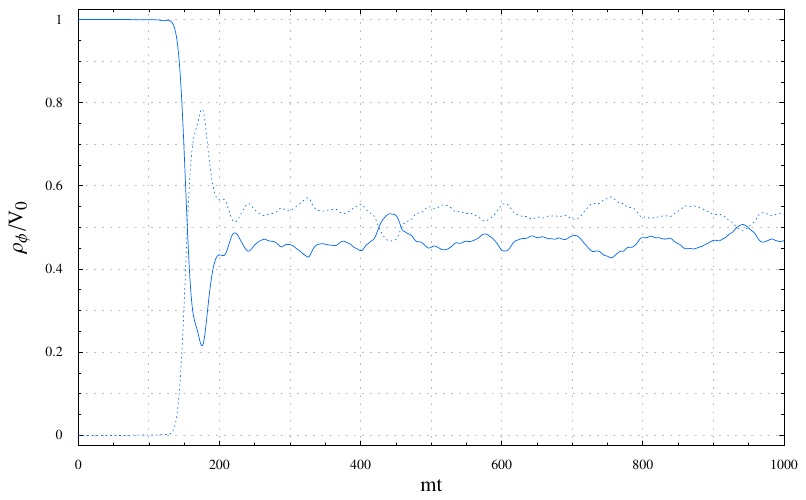}
\caption{ \label{radialangularenergy}
Radial (solid) and angular (dotted) energy densities of the flaton averaged over the lattice as a function of time ($CP-$): $\rho_\phi / V_0$ versus $mt$.
}
\end{figure}

\begin{figure}[p]
\centering
\includegraphics[width=0.95\textwidth,bb=91 3 322 146]{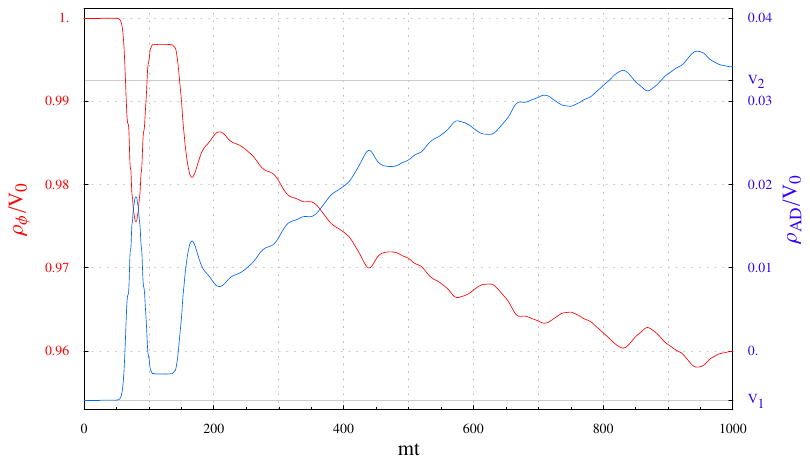}
\caption{ \label{flatonADenergy}
{\color{blue} Flaton} and {\color{red} AD} energy densities averaged over the lattice as a function of time ($CP-$): $\rho_\phi / V_0$, $\rho_\mathrm{AD} / V_0$ versus $mt$.
}
\end{figure}

The energy densities in the AD sector are shown in Figure~\ref{ADKGPenergy}.
Initially $l$ is sitting in its thermal inflation minimum and $h_d$ is at the origin, hence the potential energy is $-V_1$ and the kinetic and gradient energies are negligible.
Then the AD potential gets lifted up by the flaton and the AD fields roll in towards, and oscillate homogeneously about, the origin with negligible gradient energy.
As the flaton returns near the origin, the AD potential is dropped and $l$ returns to its thermal inflation minimum, though now oscillating with substantial kinetic energy.
The flaton then rolls out again and settles around its minimum, permanently lifting the AD potential.
The AD fields once again roll in towards, and oscillate about, the origin, though this time with rapidly growing gradient energy.
Finally, the AD energies all gradually increase, though at a decreasing rate, due to energy transfer from the flaton sector.

Figure~\ref{radialangularenergy} shows the energy transfer between the radial and angular flaton sectors.
The flaton energy is initially all in the radial component but quickly becomes equally distributed between the radial and angular components.
Over much longer timescales, far beyond the scope of this simulation, the Hubble expansion, and possibly also preheating and decay, will reduce the amplitude of the flaton oscillations about its minimum.
Then we expect the radial flaton and the axion to decouple and the energy in the axion sector to red-shift away, leaving the universe dominated by the radial flaton until it finally decays.
However, the details of this process deserve further investigation.

Figure~\ref{flatonADenergy} shows the energy transfer from the flaton sector to the AD sector.
Since the flaton sector dominates, $V_0 \gg V_1, V_2$, the energy transfer is negligible for the flaton but is significant for the AD sector.
The energy transfer could be dangerous for the conservation of the lepton number if it increases the amplitude of the $l$ field too much.
Whether this happens depends on the efficiency of the energy transfer from the flaton sector to the AD sector, which from Figure~\ref{flatonADenergy} seems to be rapidly declining, the energy flow within the AD sector to higher $k$ modes, which from Figure~\ref{spectra} seems to be very efficient but is artificially cut off at an early stage in our simulation due to the finite lattice size, and the transfer of energy from the AD sector to other sectors via thermal friction \cite{Felder:2007iz}, which is beyond the scope of this paper.
However, as the lepton is well conserved in our simulation, see Figure~\ref{leptonnumber}, despite artificially cutting off the preheating and not including thermal friction at all, we are confident it will be conserved in reality.

Thus, despite the dynamics being complex, our simulation suggests that our baryogenesis mechanism does work in this model.

\section{Conclusion}
\label{conclusion}

In this paper, we proposed a cosmological model based on a simple extension of the MSSM with superpotential
\begin{equation}
W = \lambda_u Q H_u \bar{u} + \lambda_d Q H_d \bar{d} + \lambda_e L H_d \bar{e}
+ \lambda_\mu \phi^2 H_u H_d + \frac{1}{2} \lambda_\nu \left( L H_u \right)^2
+ \lambda_\chi \phi \chi \bar\chi
\end{equation}
and key parameter condition \eq{key}.
The model is obtained from our original thermal inflation and baryogenesis model of Ref.~\cite{Jeong:2004hy} by removing the flaton self-interaction term $\lambda_\phi \phi^4$, with the flaton now stabilized by the renormalization group running of its potential.
Removing the flaton self-interaction gives the model a PQ symmetry and axion, with the scales required for thermal inflation and axions well matched.
As shown in Section~\ref{numerical}, our baryogenesis scenario works well in this model, as it did in our original model \cite{Felder:2007iz}, but now in an even more minimal setting, using essentially all the terms in the potential.

\subsubsection*{Comparison with multi-field thermal inflation axion models}

Our model, though simpler, has much in common with the multi-field thermal inflation axion models studied in Refs.~\cite{Choi:1996vz,Lazarides:2000em,Chun:2000jr,Chun:2000jx}, and our baryogenesis scenario would probably work in those models too, but there are at least two important differences.
First, in the absence of the flaton self-interaction, the flatino/axino becomes light making it the LSP, see \eq{axinomass}.
Second, the renormalization group running stabilization of the flaton's potential suppresses the flaton's mass squared around the minimum of its potential, see \eq{flatonmass}.

An axino LSP would be over-produced in the standard hot Big Bang cosmology, but in our model the flaton decays late at a temperature of order $1 \GeV$.
The lack of a flaton self-interaction suppresses direct axino production in the flaton decay, but axinos are still typically over-produced either in the flaton decay or by the decay of NLSPs in the thermal bath.
This provides the strongest constraint on our model, forcing the axino mass to be at the lower end of its expected range, $m_\axino \sim 1 \GeV$.
Avoiding over-production of axinos also requires $f_a \gtrsim 10^{11} \GeV$, which is nonetheless nicely consistent with the scale required for thermal inflation, our baryogenesis scenario and axion dark matter, see Section~\ref{constraints}.

The suppressed flaton mass suppresses the flaton decay rate to hot axions, whose over-production would conflict with Big Bang nucleosynthesis, and is the strongest constraint on the multi-field thermal inflation axion models \cite{Choi:1996vz,Lazarides:2000em,Chun:2000jr,Chun:2000jx}.
It also tends to put the flaton mass below threshold for Higgs production, reducing the decay temperature and helping avoid over-production of dark matter.

\subsubsection*{Observational tests and signatures}

Our model has a variety of observational tests and signatures.
Thermal inflation wipes out any gravitational waves at solar system scales generated during primordial inflation \cite{Mendes:1998gr}, so observation of such would rule out thermal inflation.
However, the first order phase transition that ends thermal inflation generates its own gravitational waves \cite{Easther:2008sx}.
These may be observable at future gravitational wave detectors such as BBO and DECIGO, and a correlated analysis of ultimate-DECIGO \cite{Kudoh:2005as} may probe the heart of the thermal inflation parameter space.

Thermal inflation and the following period of flaton domination redshift primordial perturbations by $10$ to $15$ $e$-folds compared with a standard hot Big Bang history.
This may have an observable effect on the primordial density perturbations.
For example, many simple models of primordial inflation predict a spectral index $n-1 = - \alpha/N + \order{1/N^2}$, where $N$ is the number of $e$-folds between horizon exit and the end of the primordial inflation.
Thermal inflation would reduce $N$ by $10$ to $15$.
One could also try to reconstruct $N$.
For example, in the simple class of models above, $(n-1)^2/n' = -\alpha$ determines the model, $(n-1)/n' = N$ determines the number of $e$-folds, and $(n-1)n''/(n')^2 = 2$ provides a check of assumptions, where $n' \equiv dn/d\ln k$, etc.

The key parameter condition for our baryogenesis scenario, \eq{key}, can be tested at future accelerators.
The roll away of the AD field may extend the thermal inflation by $5$ or $6$ $e$-folds, see \eqs{dilution}{efolds}.
The $\chi$ and $\bar\chi$ fields, needed to couple the flaton to the thermal bath, acquire intermediate scale masses after thermal inflation and, assuming that they are not all MSSM singlets, will affect the renormalisation of the MSSM couplings.
The late decay of the flaton may dilute, or even enhance, the dark matter abundance compared with a standard hot Big Bang history.

Thermal inflation axion models can be tested by the effect of the hot axions on Big Bang nucleosynthesis.
Significant parts of the parameter space of multi-field thermal inflation axion models are already ruled out by this test \cite{Choi:1996vz,Lazarides:2000em,Chun:2000jr,Chun:2000jx}.
As discussed in Sections~\ref{abr} and \ref{hotaxion}, and above, the hot axion production is suppressed in our model but is still expected to be greater than the usual thermal production, so this might also provide a test of our model in the future.

Our model, and Moxhay and Yamamoto's original single-field flaton axion model \cite{Moxhay:1984am}, can be tested by its prediction that the LSP is the axino/flatino.
Furthermore, in our model we expect $m_\axino \sim 1 \GeV$, see Sections~\ref{lsp} and \ref{flatonaxinos}.
This provides an important test of our model at future accelerators \cite{Martin:2000eq,Covi:2004rb,Brandenburg:2005he,Chun:2008rp}.
Also, the tightness of the axino bounds, see Figures~\ref{fig:con1} and \ref{fig:con2}, suggests that the dark matter should be composed of significant amounts of both axions and axinos.

\subsubsection*{Summary}

In summary, our simple and natural extension of the MSSM leads to a rich but remarkably consistent cosmology combining thermal inflation, baryogenesis, axions and axinos, with observational implications for primordial perturbations, gravitational waves, Big Bang nucleosynthesis, dark matter and particle accelerators.

\subsection*{Acknowledgements}

WIP thanks Kiwoon Choi and Yeong Gyun Kim for useful discussions, and we thank Kenji Kadota for helpful comments on an early draft of this paper.
This work was supported in part by KISTI (Korea Institute of Science and Technology Information) under the Strategic Supercomputing Support Program.
The use of the computing system of the Supercomputing Center is also greatly appreciated.
This work also used a high performance cluster that was built with funding from the Korea Astronomy and Space Science Institute (KASI) and the Astrophysical Research Center for the Structure and Evolution of the Cosmos (ARCSEC) of the Korea Science and Engineering Foundation (KOSEF).
This work was also supported by
the Korea Science and Engineering Foundation grant R01-2005-000-10404-0 funded by the Korean government (MOST),
the Korea Research Foundation grants KRF-2005-210-C000006 and KRF-2007-000-C00164 funded by the Korean Government (MOEHRD),
BK21 and ARCSEC (KOSEF).

\appendix

\section{Flaton decay rates}
\label{appendix}

\subsection{Decay to axions}

Decomposing the flaton as in \eq{flatondecomp}, the flaton kinetic term generates the radial flaton axion interaction
\begin{equation}
|\partial\phi|^2 \to \frac{\mathinner{\delta r} (\partial a)^2}{\sqrt{2}\, \phi_0}
\end{equation}
Since $m_a \simeq 0$, we get the flaton decay rate to axions
\begin{equation} \label{app:gammaphia}
\Gamma_a = \frac{m_\mathrm{PQ}^3}{64\pi \phi_0^2}
\end{equation}

\subsection{Decay via SM Higgs}

The superpotential term $\lambda_\mu \phi^2 H_u H_d$ generates the radial flaton SM Higgs interaction
\begin{equation}
\left.
\begin{array}{rcl}
\displaystyle
|\lambda_\mu|^2 |\phi|^4 \left( |H_u|^2 + |H_d|^2 \right)
& \to &
\displaystyle
\sqrt{2}\, \frac{|\mu|^2}{\phi_0} \mathinner{\delta r} h^2
\\[3ex]
\displaystyle
B \lambda_\mu \phi^2 H_u H_d + \textrm{c.c.}
& \to &
\displaystyle
- \sqrt{2}\, \frac{|B|^2 |\mu|^2}{m_A^2 \phi_0} \mathinner{\delta r} h^2
\end{array}
\right\} \to \sqrt{2} \left( 1 - \frac{ |B|^2}{m_A^2} \right) \frac{|\mu|^2}{\phi_0} \mathinner{\delta r} h^2
\end{equation}
where
\begin{equation}
m_A^2 = - m_{H_u}^2 + m_{H_d}^2 + 2 |\mu|^2
\end{equation}
and we have used \eq{mu}.
Therefore the flaton decay rate via (possibly virtual) SM Higgses is \cite{Grau:1990uu}
\begin{eqnarray}
\Gamma_{\phi \to hh} & = & A
\int_0^{m_\mathrm{PQ}} \frac{2p \d{p} m_h \Gamma_h}{\pi \left[ \left( p^2 - m_h^2 \right)^2 + \left( m_h \Gamma_h \right)^2 \right]}
\int_0^{m_\mathrm{PQ} - p} \frac{2q \d{q} m_h \Gamma_h}{\pi \left[ \left( q^2 - m_h^2 \right)^2 + \left( m_h \Gamma_h \right)^2 \right]}
\nonumber \\ && \times
\sqrt{ 1 - \frac{2 p^2}{m_\mathrm{PQ}^2} - \frac{2 q^2}{m_\mathrm{PQ}^2} - \frac{2 p^2 q^2}{m_\mathrm{PQ}^4} + \frac{p^4}{m_\mathrm{PQ}^4} + \frac{q^4}{m_\mathrm{PQ}^4} }
\end{eqnarray}
where
\begin{equation} \label{A}
A = \frac{|\mu|^4}{4\pi m_\mathrm{PQ} \phi_0^2} \left( 1 - \frac{|B|^2}{m_A^2} \right)^2
\end{equation}
and $\Gamma_h$ is the decay rate of the SM Higgs.
The dominant decay mode of the SM Higgs is to bottom quarks, giving \cite{Yao:2006px}
\begin{equation} \label{littlegammah}
\gamma_h \equiv \frac{\Gamma_h}{m_h} \simeq \frac{3 g_2^2 m_b^2}{32\pi m_W^2} \left( 1 - \frac{4 m_b^2}{m_h^2} \right)^{3/2} \sim 10^{-5}
\end{equation}
where $g_2 \simeq 0.65$ is the $\mathrm{SU}(2)$ gauge coupling, $m_b \simeq 5 \GeV$ is the bottom quark mass and $m_W \simeq 80 \GeV$ is the $W$ boson mass.
Therefore, defining $x = p / m_\mathrm{PQ}$, $y = q / m_\mathrm{PQ}$ and $\xi = m_h / m_\mathrm{PQ}$, we have
\begin{eqnarray} \label{gammaphihoverA}
\Gamma_{\phi \to hh}
& = & A \int_0^1 \d{x} \int_0^{1-x} \d{y} \frac{4xy \gamma_h^2 \xi^2 \sqrt{ 1 - 2 x^2 - 2 y^2 - 2 x^2 y^2 + x^4 + y^4 }}{\pi^2 \left[ \left( x^2 - \xi^2 \right)^2 + \gamma_h^2 \xi^2 \right] \left[ \left( y^2 - \xi^2 \right)^2 + \gamma_h^2 \xi^2 \right]}
\\[3ex]
& = & A \left\{
\begin{array}{lrc}
\sqrt{ 1 - 4 \xi^2 } + \order{\gamma_h}
& \textrm{for} & \xi < \frac{1}{2}
\\[2ex]
2 \gamma_h \int_0^{1-\xi} \frac{2x \d{x} \sqrt{ \left( 1 - \xi^2 \right)^2 - 2 \left( 1 + \xi^2 \right) x^2 + x^4 }}{\pi \left( x^2 - \xi^2 \right)^2} + \order{\gamma_h^2}
& \textrm{for} & \frac{1}{2} < \xi < 1
\\[2ex]
\gamma_h^2 \int_0^1 \d{x} \int_0^{1-x} \d{y} \frac{4xy \sqrt{ 1 - 2 x^2 - 2 y^2 - 2 x^2 y^2 + x^4 + y^4 }}{\pi^2 \left( x^2 - \xi^2 \right)^2 \left( y^2 - \xi^2 \right)^2}
 + \order{\gamma_h^3}
& \textrm{for} & 1 < \xi
\end{array}
\right. \nonumber \\ \label{gammah}
\end{eqnarray}

\subsection{Decay to $b$ and $\bar{b}$ via flaton-Higgs mixing}

The superpotential term $\lambda_\mu \phi^2 H_u H_d$ induces flaton-Higgs mixing, allowing the flaton mass eigenstate to decay directly to SM fermions.

The neutral $CP$ even Higgs states are
\begin{eqnarray}
H_u^0 & = & \Theta_{H_u^0}^h h + \Theta_{H_u^0}^H H
\\
H_d^0 & = & \Theta_{H_d^0}^h h + \Theta_{H_d^0}^H H
\end{eqnarray}
where $h$ and $H$ are the mass eigenstates, $h$ being the SM Higgs.
The mass matrix elements are
\begin{eqnarray}
\mathcal{M}^2_{H_u^0 H_u^0} & = & \frac{1}{2} m_A^2 \left( 1 + \cos{2\beta} \right) + \frac{1}{2} m_Z^2 \left( 1 - \cos{2\beta} \right)
\\
\mathcal{M}^2_{H_u^0 H_d^0} & = & - \frac{1}{2} \left( m_A^2 + m_Z^2 \right) \sin{2\beta}
\\
\mathcal{M}^2_{H_d^0 H_d^0} & = & \frac{1}{2} m_A^2 \left( 1 - \cos{2\beta} \right) + \frac{1}{2} m_Z^2 \left( 1 + \cos{2\beta} \right)
\end{eqnarray}
giving the tree level Higgs masses
\begin{eqnarray}
m_h^2 & = & \frac{1}{2} \left[ \left( m_A^2 + m_Z^2 \right) -  \sqrt{ \left( m_A^2 + m_Z^2 \right)^2 - 4 m_A^2 m_Z^2 \cos^2{2\beta} } \right]
\\
m_H^2 & = & \frac{1}{2} \left[ \left( m_A^2 + m_Z^2 \right) +  \sqrt{ \left( m_A^2 + m_Z^2 \right)^2 - 4 m_A^2 m_Z^2 \cos^2{2\beta} } \right]
\end{eqnarray}
where
\begin{eqnarray}
\label{mA}
m^2_A & = & m^2_{H_u} + m^2_{H_d} + 2 |\mu|^2
\\
\sin{2\beta} & = & \frac{2|B\mu|}{m^2_A}
\\ \label{mz}
m^2_Z & = & 2 \left( \frac{m^2_{H_d} - m^2_{H_u} \tan^2{\beta}}{\tan^2{\beta} - 1} \right) - 2 |\mu|^2
\end{eqnarray}
$A$ is the neutral $CP$ odd Higgs boson and $\tan{\beta}$ is defined as
\begin{equation}
\tan{\beta} \equiv \frac{v_u}{v_d}
\end{equation}
where $v_u$ and $v_d$ are the magnitudes of the vacuum expectation values of $H_u$ and $H_d$, and the electroweak symmetry breaking scale is
\begin{equation}
v^2 = v_u^2 + v_d^2 = (174\GeV)^2
\end{equation}

The flaton mixes with $h$ and $H$
\begin{eqnarray}
h & = & \Theta_h^{\hat{h}} \hat{h} + \Theta_h^{\hat{H}} \hat{H} + \Theta_h^{\hat\phi} \hat\phi
\\
H & = & \Theta_H^{\hat{h}} \hat{h} + \Theta_H^{\hat{H}} \hat{H} + \Theta_H^{\hat\phi} \hat\phi
\\
\phi & = & \Theta_\phi^{\hat{h}} \hat{h} + \Theta_\phi^{\hat{H}} \hat{H} + \Theta_\phi^{\hat\phi} \hat\phi
\end{eqnarray}
where $\hat{h}$, $\hat{H}$ and $\hat\phi$ are the mass eigenstates, due to the mass matrix elements
\begin{eqnarray}
\mathcal{M}^2_{\phi \phi} & = & m_\mathrm{PQ}^2
\\
\mathcal{M}^2_{H_u^0 \phi} & = & 2 \sin{\beta} \left( 2 | \mu |^2 - m_A^2 \cos^2{\beta} \right) \frac{v}{\phi_0}
\\
\mathcal{M}^2_{H_d^0 \phi} & = & 2 \cos{\beta} \left( 2 | \mu |^2 - m_A^2 \sin^2{\beta} \right) \frac{v}{\phi_0}
\end{eqnarray}
Since the decay is dominated by the bottom quark channel for $m_\phi \lesssim 1 \TeV$ \cite{Chun:2000jr}
\begin{equation}
\lambda_b H_d^0 b \bar{b} \to \lambda_b \Theta_{H_d^0}^{\hat\phi} \hat\phi b \bar{b}
\end{equation}
\begin{equation} \label{app:gammaphibb}
\Gamma_{\hat\phi \to b \bar{b}} = \frac{3}{16 \pi} \left| \lambda_b \Theta_{H_d^0}^{\hat\phi} \right|^2 m_\mathrm{PQ} \left( 1 - \frac{4 m_b^2}{m_\mathrm{PQ}^2} \right)^{3/2}
\end{equation}
we are only interested in the $\hat\phi$ $H_d^0$ mixing
\begin{equation}
\Theta_{H_d^0}^{\hat\phi} = \Theta_{H_d^0}^h \Theta_h^{\hat\phi} + \Theta_{H_d^0}^H \Theta_H^{\hat\phi}
\end{equation}
Diagonalizing the Higgs mass matrix gives
\begin{eqnarray}
\Theta_{H_d^0}^h
& = & \frac{\mathcal{M}_{H_u^0 H_d^0}^2}{\sqrt{\left( \mathcal{M}_{H_d^0 H_d^0}^2 - m_h^2 \right)^2 + \mathcal{M}_{H_u^0 H_d^0}^4}}
\\
& = & - \frac{A_\Theta \cos{\beta}}{\sqrt{A_\Theta^2 + 2 A_\Theta B_\Theta \sin{\beta} + B_\Theta^2}}
\\
\Theta_{H_d^0}^H
& = & \frac{\mathcal{M}_{H_d^0 H_d^0}^2 - m_h^2}{\sqrt{\left( \mathcal{M}_{H_d^0 H_d^0}^2 - m_h^2 \right)^2 + \mathcal{M}_{H_u^0 H_d^0}^4}}
\\
& = & \frac{A_\Theta \sin{\beta} + B_\Theta}{\sqrt{A_\Theta^2 + 2 A_\Theta B_\Theta \sin{\beta} + B_\Theta^2}}
\end{eqnarray}
and perturbatively diagonalizing the flaton Higgs mass matrix gives
\begin{eqnarray}
\Theta_h^{\hat\phi}
& = & - \frac{\mathcal{M}_{H_u^0 H_d^0}^2 \mathcal{M}_{H_d^0 \phi}^2 - \left( \mathcal{M}_{H_d^0 H_d^0}^2 - m_h^2 \right) \mathcal{M}_{H_u^0 \phi}^2}{\left( m_h^2 - \mathcal{M}_{\phi \phi}^2 \right) \sqrt{\left( \mathcal{M}_{H_d^0 H_d^0}^2 - m_h^2 \right)^2 + \mathcal{M}_{H_u^0 H_d^0}^4}}
\\
& = & \frac{ 4 \left( 1 - \frac{|B|^2}{m_A^2} \right) |\mu|^2 v}{\left( m_h^2 - m_\mathrm{PQ}^2 \right) \phi_0} \left[ \frac{A_\Theta + B_\Theta \left( \frac{4 | \mu |^2 - 2 m_A^2 \cos^2{\beta}}{4 | \mu |^2 - m_A^2 \sin^2{2\beta}} \right) \sin{\beta}}{\sqrt{A_\Theta^2 + 2 A_\Theta B_\Theta \sin{\beta} + B_\Theta^2}} \right]
\\
\Theta_H^{\hat\phi}
& = & - \frac{\left( \mathcal{M}_{H_d^0 H_d^0}^2 - m_h^2 \right) \mathcal{M}_{H_d^0 \phi}^2 + \mathcal{M}_{H_u^0 H_d^0}^2 \mathcal{M}_{H_u^0 \phi}^2}{\left( m_H^2 - \mathcal{M}_{\phi \phi}^2 \right) \sqrt{\left( \mathcal{M}_{H_d^0 H_d^0}^2 - m_h^2 \right)^2 + \mathcal{M}_{H_u^0 H_d^0}^4}}
\\
& = & - \frac{m_A^2 v \sin{4\beta}}{2 \left( m_H^2 - m_\mathrm{PQ}^2 \right) \phi_0} \left[ \frac{A_\Theta + B_\Theta \left( \frac{2 | \mu |^2 - m_A^2 \sin^2{\beta}}{m_A^2 \sin{\beta} \cos{2\beta}} \right)}{\sqrt{A_\Theta^2 + 2 A_\Theta B_\Theta \sin{\beta} + B_\Theta^2}} \right]
\end{eqnarray}
where
\begin{eqnarray}
A_\Theta & = & \left( m_A^2 + m_Z^2 \right) \sin{\beta}
\\
B_\Theta & = & m_Z^2 \cos{2\beta} - m_h^2
\end{eqnarray}
Thus the mixing coefficient $\Theta_{H_d^0}^{\hat\phi}$ has a complicated dependence on the parameters, but can be roughly parameterized as
\begin{eqnarray}
\Theta_{H_d^0}^{\hat\phi} & \sim & 4 C_\Theta \left( 1 - \frac{|B|^2}{m_A^2} \right) \frac{|\mu|^2 v \cos{\beta}}{\left( m_\mathrm{PQ}^2 - m_h^2 \right) \phi_0}
\end{eqnarray}
with $C_\Theta \sim 1$.
With this parameterization, \eq{app:gammaphibb} becomes
\begin{equation} \label{gammaphiff}
\Gamma_{\hat\phi \to b \bar{b}} \sim 12 C_\Theta^2 A \left( 1 - \frac{4 m_b^2}{m_\mathrm{PQ}^2} \right)^\frac{3}{2} \left( \frac{m_b^2}{m_\mathrm{PQ}^2} \right) \left| \frac{m_\mathrm{PQ}^2}{m_\mathrm{PQ}^2 - m_h^2} \right|^2
\end{equation}
where $A$ is defined in \eq{A}.

\subsection{Decay to gluons via $\chi$ and $\bar\chi$}

The superpotential term $\lambda_\chi \phi \chi \bar\chi$ allows the flaton to decay to SM gauge fields via a virtual $\chi \bar\chi$ loop.
The dominant decay channel is via heavy quark superfields to gluons \cite{Chun:2000jr}
\begin{equation} \label{gammaphigg}
\Gamma_{\phi \to g g}
\simeq \frac{\alpha_3^2 N_q^2 m_\mathrm{PQ}^3}{144\pi^3 \phi_0^2} \left( 1 + \frac{95 \alpha_3}{4\pi} \right)
\end{equation}
where $\alpha_3 \equiv g_3^2 / (4\pi) \simeq 0.1$.

\subsection{Decay to axinos}
\label{flatonaxinodecay}

Axino mass is generated at one-loop and given by \eq{axinomass}
\begin{equation}
m_\axino = \frac{1}{16 \pi^2} \sum_\chi \lambda_\chi^2 A_\chi
\end{equation}
The $|\phi|$ dependent renormalisation of $\lambda_\chi$ and $A_\chi$ is \cite{Martin:1997ns}
\begin{eqnarray}
\frac{d\lambda_\chi^2}{d\ln|\phi|} & = & \frac{\lambda_\chi^2}{8\pi^2}
\left( \sum_{\chi'} \left|\lambda_{\chi'}\right|^2 + 2 \left|\lambda_\chi\right|^2 - 4 \sum_a \fn{C_a}{\chi} g_a^2 \right)
\\
\frac{d A_\chi}{d\ln|\phi|} & = & \frac{1}{8\pi^2}
\left( \sum_{\chi'} \left|\lambda_{\chi'}\right|^2 A_{\chi'} + 2 \left|\lambda_\chi\right|^2 A_\chi + 4 \sum_a \fn{C_a}{\chi} g_a^2 M_a \right)
\end{eqnarray}
where the $\fn{C_a}{\chi}$ are the quadratic Casimir invariants, for example $\fn{C_3}{\chi} = 4/3$ for a quark superfield $\chi$.
This generates the coupling of the flaton to the axino, given in \eq{flatonaxinocoupling}, with
\begin{eqnarray}
\alpha_\axino & = & \frac{d \ln m_\axino}{d\ln|\phi|}
\\
& = & \frac{
\sum_\chi \lambda_\chi^2
\left[ \sum_{\chi'} \left|\lambda_{\chi'}\right|^2 A_{\chi'}
+ \sum_{\chi'} \left|\lambda_{\chi'}\right|^2 A_\chi
+ 4 \left|\lambda_\chi\right|^2 A_\chi + 4 \sum_a \fn{C_a}{\chi} g_a^2 \left( M_a - A_\chi \right)
\right]
}{8\pi^2 \sum_\chi \lambda_\chi^2 A_\chi}
\nonumber \\
\end{eqnarray}

\end{document}